\documentclass[12pt]{article}
\usepackage[active]{srcltx}
\usepackage{amsbsy}
\usepackage{amsfonts}
\usepackage{amssymb}
\usepackage{amsmath}
\usepackage{graphicx}
\usepackage{subfigure}
\usepackage[usenames, dvipsnames]{color}

\makeatletter
\renewcommand\section{\@startsection {section}{1}{\z@}%
                                   {-3.5ex \@plus -1ex \@minus -.2ex}%
                                   {2.3ex \@plus.2ex}%
                                   {\normalfont\large\bfseries}}
\renewcommand\subsection{\@startsection{subsection}{2}{\z@}%
                                   {-3.25ex\@plus -1ex \@minus -.2ex}%
                                   {1.5ex \@plus .2ex}%
                                   {\normalfont\normalsize\bfseries}}
\makeatother

\def\IR{\mathbb{R}}

\def\be{\begin{eqnarray}}
\def\ee{\end{eqnarray}}
\def\Tr{{\rm Tr}}

\def\quart{{\textstyle{\frac{1}{4}}}}
\def\half{{\textstyle{\frac{1}{2}}}}
\def\bfbeta{\beta}
\def\h{{\tilde h}}

\def\NeqFour{{\cal N}=4}
\def\NeqOne{{\cal N}=1}

\def\C{{\rm C}}
\def\U{{\rm U}}
\def\E{{\rm E}}
\def\B{{\rm B}}

\def\q{{\rm q}}
\def\gym{g_{_{YM}}}

\begin{document}

\vspace{ -3cm}
\thispagestyle{empty}
\vspace{-1cm}

\rightline{}

\begin{center}
\vspace{1cm}
{\Large\bf

On the non-planar $\boldsymbol{\beta}$-deformed $\NeqFour$ super-Yang-Mills theory

\vspace{1.2cm}

   }

\vspace{.2cm}
 {Q.~Jin\footnote{qxj103@psu.edu}
and
  R.~Roiban\footnote{radu@phys.psu.edu}}
\\

\vskip 0.6cm

{\em

Department of Physics, The Pennsylvania  State University,\\
University Park, PA 16802, USA

 }

\vspace{.2cm}

\end{center}

\begin{abstract}

The $\beta$-deformation is one of the two superconformal deformations of the 
$\NeqFour$ super-Yang-Mills theory. At the planar level it shares all of its properties 
except for supersymmetry, which is broken to the minimal amount. The tree-level 
amplitudes of this theory exhibit new features which depart from the commonly 
assumed properties of gauge theories with fields in the adjoint representation.

We analyze in detail complete one-amplitudes and a nonplanar two-loop amplitude of this theory and show 
that, despite having only $\NeqOne$ supersymmetry, the two-loop amplitudes
have a further-improved ultraviolet behavior. This phenomenon is a counterpart of
a similar improvement previously observed in the double-trace amplitude of the 
$\NeqFour$ super-Yang-Mills theory at three and four loop order and points to the 
existence of additional structure in both the deformed and undeformed theories.

\end{abstract}

\newpage

\section{Introduction}

The maximally supersymmetric Yang-Mills (sYM) theory has provided a valuable arena for 
devising new and powerful methods for perturbative quantum field theory computations. 
In the planar limit, this effort exposed remarkable properties of the theory, such as dual 
(super)conformal invariance and integrability, as well as unexpected relations between {\it a 
priori} unrelated quantities, such as the relation between certain null polygonal Wilson loops, 
scattering amplitudes and correlation functions of gauge invariant operators. Each of these 
presentations of scattering amplitudes, as a Wilson loop or as a (square root of a) correlation
function, manifestly realizes one of the two conformal symmetries of the theory.
It has been suggested that, in four dimensions, the symmetries of the theory together with 
the leading singularities of amplitudes completely determine 
the planar loop integrand to all orders in perturbation theory \cite{ArkaniHamed:2010kv}.

At the nonplanar level the symmetry structure of the theory is expected to be different. 
While comparatively less studied, explicit calculations of complete amplitudes 
\cite{Bern:2008pv, Bern:2010tq} expose unexpected structure:
the color/kinematics duality and the BCJ relations \cite{Bern:2008qj}, generalizing the $U(1)$ decoupling 
identities and the Kleiss-Kuijff relations, provide a direct link between the planar and non-planar 
parts of amplitudes. They also suggest that the kinematic numerators of integrands of amplitudes 
(once organized in a specific way) obey an algebra akin to that of the color factors. The 
fundamental origin as well as its implications and limitations are not completely understood 
(see however \cite{Monteiro:2011pc} for an interesting discussion in the self-dual sector).
The color/kinematic duality has also been discussed in less supersymmetric theories~\cite{Bern:2008qj}.

A possible strategy to identify the essential elements underlying the fascinating properties 
of the $\NeqFour$ sYM theory at the non-planar level is to deform this theory in a way that 
does not alter its planar properties and then analyze the resulting theory. Since such 
deformations necessarily break supersymmetry, this analysis would also probe, in a more 
general context, other aspects of the undeformed theory: 
the possibility and subtleties associated to using the 
Coulomb branch as infrared regulator even at the planar level; the consistency conditions on 
massless S-matrices discussed in \cite{Benincasa:2007xk}, etc. 
A further appeal of this strategy is that, by reducing the number 
of supercharges, unexpected phenomena unrelated to maximal supersymmetry but  
hidden by it may emerge at low loop order. 

The two deformations of the $\NeqFour$ sYM theory which preserve conformal 
invariance to all orders in perturbation theory at both planar and non-planar level 
have been identified by Leigh and Strassler in~\cite{Leigh:1995ep}. In both cases the arguments described 
there show that there exists a relation between the gauge coupling, the strength of 
the deformation and the number of colors which, if satisfied, guarantees conformal  
invariance of the theory. 

The so-called $\beta$-deformation was extensively studied from several standpoints.
The planar scattering amplitudes have been discussed in \cite{Khoze:2005nd}
where it was shown that, for real $\beta$, they are inherited from the undeformed theory 
and thus are dual conformally invariant.
From the current perspective this result reinforces the link between integrability 
and dual conformal invariance. Indeed,  the $\beta$-deformation preserves integrability
both at weak \cite{Beisert:2005if} and strong coupling \cite{Frolov:2005dj} and
the construction of  its string theory dual \cite{Lunin:2005jy, Frolov:2005dj} 
suggests that the arguments of  \cite{Berkovits:2008ic, Beisert:2008iq} relating 
integrability and dual conformal symmetry should apply in this case as well.

While the existence of a line of fixed points is guaranteed by the arguments of 
\cite{Leigh:1995ep}, the precise relation between parameters is known in general 
only through two-loop order\footnote{In the planar limit this relation is known to all
loop orders. There have been in fact two definitions of finiteness, which provide 
slightly different constraints on parameters. One of them demands that all beta-functions 
vanish while the other demands that all correlation functions are finite.}. Our calculations
of four-point amplitudes at one- and two-loop level will confirm the known constraint. 
We will argue that this constraint must be modified at three loops. This is similar in spirit 
to the argument of \cite{Khoze:2005nd} that, for complex deformation parameter, the 
planar relation between the deformed theory and $\NeqFour$ sYM theory breaks down at 
five loops. 
%

As all massless theories, the $\beta$-deformed $\NeqFour$ sYM theory has infrared 
divergences and therefore requires regularization. While in gauge theories IR divergences are 
conventionally dimensionally regulated, several possible IR regularizations are available. 
In the planar undeformed theory it has been argued \cite{Alday:2009zm} that moving off the 
origin of the Coulomb branch and partly breaking the gauge symmetry as $SU(N+M)\mapsto 
SU(N)\times SU(M)$  will on the one hand regularize IR divergences while on 
the other will preserve dual conformal symmetry if masses of fields (or vacuum expectation 
values of scalars) are assigned suitable transformation rules. 
An essential ingredient in these arguments is that, as a consequence of extended 
supersymmetry, the relation between vacuum expectation values and field masses does 
not receive quantum corrections. 
The minimal supersymmetry of the $\beta$-deformed theory does not guarantee the absence 
of such corrections which, as we will see, appear at the very least due to the corrections to the
conformality condition. Therefore, using the Higgs regularization in the $\beta$-deformed theory 
at higher loops requires a certain amount of care as one should isolate the quantum 
corrections to the regulator. For the same reason, the close relation between the $\NeqFour$ 
and the $\beta$-deformed planar amplitudes \cite{Khoze:2005nd}, while present in dimensional
regularization, does not appear to hold in a straightforward way in the Higgs regularization.

By completely breaking the gauge symmetry, $SU(N)\rightarrow U(1)^{N-1}$, the Higgs 
regularization may also be used at nonplanar level both in the $\beta$-deformed and in
the undeformed theory.
While we will draw information on the structure of amplitudes in the presence of such a 
regulator, we will carry out calculations in dimensional regularization.
It is interesting to note that, unlike in the $\NeqFour$ theory, dimensional regularization
changes the total number of degrees of freedom of the theory. Indeed, as an $\NeqOne$ 
theory that does not have an ${\cal N}=2$ extension, IR dimensional regularization will
dimensionally-continue vector multiplets without reducing the number of chiral multiplets. 
For this reason the notion of critical dimension as the dimension in which the first logarithmic 
divergence appears at some fixed loop order is not a well-defined concept in the 
$\beta$-deformed theory. 
One may nevertheless formally continue to arbitrary dimension the integrand of an 
amplitude  and thus define a quantity which, as we will see, captures the convergence 
properties of the integrand and (discontinuously) becomes the standard critical dimension 
as the deformation parameter is set to zero.

We will begin in the next section with a brief review of the $\beta$-deformed sYM theory, 
discuss its Coulomb branch and outline the main differences between its scattering amplitudes 
and those of the undeformed theory. In \S~\ref{general_loops} we will discuss the general
features of loop calculations in this theory through the generalized unitarity method,
the constraints imposed by supersymmetry and describe non-planar all-loop results that are 
inherited from the undeformed theory.
In \S~\ref{oneloop} we present the explicit expressions for representatives of the five 
different classes of four-point amplitudes and discuss the properties of their infrared 
divergences.
In \S~\ref{twoloop} we describe the first nontrivial two-loop amplitude which is sensitive 
to the deformation parameter and analyze its properties. We will see that, unlike the 
one-loop amplitudes which exhibit the standard properties of a finite $\NeqOne$ theory,
the two-loop amplitudes we evaluate have better UV convergence properties than 
one might expect based on  $\NeqOne$ supersymmetry alone.
We summarize our results in \S~\ref{conclusion} and speculate on the relation
of the improved UV behavior  and the undeformed theory.

\section{The $\beta$-deformed $\NeqFour$ sYM theory and its features \label{review}}

The $\beta$-deformed $\mathcal{N}$=4 theory is one of the two ${\cal N}=1$ exactly marginal 
deformations\footnote{The second exactly marginal deformation is given by $\delta W = \Tr[\Phi_1^3+
\Phi_2^3+\Phi_3^3]$.} of the maximally supersymmetric Yang-Mills theory in four dimensions 
\cite{Leigh:1995ep}; it is obtained by adding to the superpotential of the ${\cal N}=4$ sYM 
theory the term  ${\delta W}=\Tr[\Phi^1\{\Phi^2, \Phi^3\}]$. 
The action may be written in $\mathcal{N}$=1 superspace as
\begin{eqnarray}
\label{action}
S&=&\int d^{4}xd^{4}\theta\, \Tr[e^{-\gym V}\bar{\Phi}_{i}e^{\gym V}\Phi^{i}]
\\
&+&\frac{1}{2\gym ^{2}}\int d^{4}xd^{2}\theta \Tr[W^{\alpha}W_{\alpha}]
+ih\int d^{4}xd^{2}\theta\, \Tr[q\Phi^{1}\Phi^{2}\Phi^{3}-q^{-1}\Phi^{1}\Phi^{3}\Phi^{2}]+{\rm h.c.} \ .
\nonumber
\end{eqnarray}
This superspace, inherited from the one used for the $\NeqFour$ theory, manifestly preserves 
the fourth component of the $\NeqFour$ quartet of supercharges. The parameter $q$ is
customarily parametrized as $q=\exp(i\beta)$ with $\beta$ being either real or complex. 
The two choices have distinct quantum mechanical properties with only the former theory, with
$\beta\in\IR$, being integrable at the planar level. 
%
%
In the following we will assume that $\beta$ is real.

It has been pointed out in \cite{Lunin:2005jy} that the addition of $\delta W$ may be interpreted 
as a noncommutative deformation of $\NeqFour$ sYM theory in the R-symmetry directions; 
instead of space-time momenta, the relevant Moyal-like product involves the three $U(1)$ 
charges $\q=(\q_1,\q_2, \q_3)$ inherited from the $SU(4)$ symmetry of $\NeqFour$ sYM theory:
\footnote{For the present choice of superspace, the product of component fields is directly
inherited from the product of superfields because the charge vector of the manifestly-realized supercharge 
does not affect the phase in eq.~(\ref{Moyal}).}
\be
\Phi^i\Phi^j\mapsto e^{i\epsilon^{abc}\beta_a\q^i_b\q^j_c}\Phi^i\Phi^j \equiv 
e^{i\beta^{bc}\q^i_b\q^j_c}\Phi^i\Phi^j\equiv 
e^{i\q^i\wedge \q^j}\Phi^i\Phi^j\ ,
\label{Moyal}
\ee
where $\beta_1=\beta_2=\beta_3=\beta$. It is not difficult to see that such a phase becomes 
trivial whenever at least one of the two fields belongs to the ${\cal N}=1$ vector multiplet. 
In the following we will  denote by $[\bullet, \bullet]_\beta$ the commutator in which the 
product of the two entries has been replaced by the Moyal product  (\ref{Moyal}), {\it i.e.}
\be
[A,B]_\beta = e^{i\beta^{bc}\q^A_b\q^A_c}AB-e^{i\beta^{bc}\q^B_b\q^A_c}BA \ .
\ee

The action (\ref{action}) exhibits three global U(1) symmetries. One of them is the
$U(1)$ R-symmetry of the $\NeqOne$ superspace. The other two simultaneously rephase 
two of the three chiral superfields as:
\be
(\Phi^1,\Phi^2,\Phi^3)\mapsto(\Phi^1,e^{i\alpha_1}\Phi^2,e^{-i\alpha_1}\Phi^3)
\qquad
(\Phi^1,\Phi^2,\Phi^3)\mapsto(e^{i\alpha_2}\Phi^1,e^{-i\alpha_2}\Phi^2,\Phi^3) \ .
\ee
Three linear combinations of these symmetries form the Cartan subalgebra of the $SU(4)$ 
R-symmetry of $\NeqFour$ sYM theory under which fermions transform in the fundamental 
representation (spinor representation of $SO(6)$) with charges
\be
\q^1=(\half, -\half, -\half)
~,~~
\q^2=(-\half, \half, -\half)
~,~~
\q^3=(-\half, -\half, \half)
~,~~
\q^4=(\half, \half, \half) \ ;
\ee
the index $"4"$ labels the supercharges preserved by the $\NeqOne$ superspace 
in eq.~(\ref{action}).
The scalar fields transform in the two-index antisymmetric representation (vector representation 
of $SO(6)$) and thus their charges may be obtained in terms of those of fermions as
\be
\phi^{ij} \ ,~~\q^{ij}=\q^i+\q^j
\quad,\qquad
\phi^{k4}\equiv{\overline{\phi^{ij}}} \ ,~~\q^{k4}=\q^k+\q^4=-\q^i-\q^j 
~,~~i,j,k=1,2,3\ .
%
\ee
This is the same labeling used in the $\NeqFour$ on-shell superspace. The phase associated
to a product of fields with charges $\q_1,\dots, \q_n$ is just
\be
\label{phase_general}
\phi_1\dots \phi_n \mapsto e^{i\varphi(\q(\phi_1),\dots,\q(\phi_n))}\phi_1\dots \phi_n
\qquad
\varphi(\q(\phi_1),\dots,\q(\phi_n)) = \sum_{i<j=1}^n\q(\phi_i)\wedge \q(\phi_j) \ .
\ee
This expression may be further simplified if the charge vectors have additional properties, such 
as a vanishing total charge $\sum_{i=1}^n\q(\phi_i)=0$.

%
%
Based on the properties of noncommutative Feynman graphs 
\cite{Filk:1996dm} it has been argued that, in dimensional regularization and for real $\beta$, 
all planar scattering amplitudes of the $\beta$-deformed $\NeqFour$ theory are the same -- up to 
constant $\beta$-dependent  phases -- as the scattering amplitudes. For complex $\beta$ this 
equivalence appears to break down at five loop order \cite{Khoze:2005nd}; it was moreover shown 
\cite{Elmetti:2006gr} that finiteness of the planar propagator corrections require that $\beta$ be 
real.
 
The deformed theory distinguishes between an $U(N)$ and an $SU(N)$
gauge group. In the former case, the scalar fields valued in the diagonal $U(1)$ factor 
do not decouple and their coupling constant flows for all values of the parameters of the 
theory. They decouple in the infrared, where their coupling constants reach zero and the Lagrangian
becomes that of the $SU(N)$ theory. In a trace-based presentation, the component Lagrangian 
for this gauge group is
\begin{eqnarray}
\label{def_lag}
 {\cal L}&=&{\cal L}_1+{\cal L}_2
 \\
{\cal L}_1  &=& \Tr \Big[ \quart F_{\mu \nu }^{~~2}+\half\bar{\psi}{}^4%
             {D}\!\!\!\!\slash\psi^4
             +\overline{D_{\mu }\phi }{}_iD_{\mu }\phi^i +\half 
             \bar{\psi}_i{D}\!\!\!\!\slash\psi^i  \nonumber\\
 {}& &\!\!\!\!\!\!\!\!\!\!\!\!\! +\sqrt{2} \gym 
({\psi}^\alpha{}^4 L[\bar{\phi}{}_i,\,\psi^i_\alpha]
  -\bar{\psi}^{\dot\alpha}_i R [\phi^i,\,\psi^4_{\dot\alpha}])   
 + \frac{i}{\sqrt2} \left(h
\epsilon_{ijk} {\psi}{}^\alpha{}^i L [\phi_b^j,\,
\psi_\alpha ^k]_ \bfbeta
	  +{\bar h}\epsilon^{ijk} \bar{\psi}{}^{\dot\alpha}_i R [\bar{\phi}_j,\, 
{\bar\psi}{}_{\dot\alpha}{}_k]_ \bfbeta\right)\nonumber    \\ 
 {}& &\!\!\!\!\!\!\!\!\!\!\!\!\!  +\half g^{2}([\bar{\phi}_i,\, \phi^i])^{2}
  -\half{|h|^2}  \epsilon_{ijk} \epsilon^{ilm}
   [\phi^j,\, \phi^k]_\bfbeta [\bar{\phi}_l,\,\bar{\phi}_m]_\bfbeta ~~ \Big]
\cr
{\cal L}_2&=&+\frac{1}{2N}|h^2|\epsilon_{ijk} \epsilon^{ilm} 
\Tr[[\phi^j,\, \phi^k]_ \bfbeta]\Tr[ [\bar{\phi}_l,\,\bar{\phi}_m]_ \bfbeta ]
\end{eqnarray}
where $L$ and $R$ are chiral projectors and ${\cal L}_2$ appears upon integrating 
out the auxiliary fields due to a tracelessness condition.
In the component Lagrangian one may further generalize the Moyal product (\ref{Moyal})
by replacing $\beta$ with a generic $3\times 3$ antisymmetric 
matrix \cite{Frolov:2005dj, Beisert:2005if}. Such a generalization completely 
breaks supersymmetry and the resulting theory appears to be unstable 
\cite{DR}.

The original arguments of Leigh and Strassler \cite{Leigh:1995ep} imply that the 
theory with an $SU(N)$ gauge group is conformally invariant if the three parameters 
$\gym$, $h$ and $q$ obey one relation, 
$\gamma(\gym,h,q)=0$.
The complete functional form of $\gamma$ is not known; through two-loop order and for  
$|q|=1$, vanishing of the $\beta$-function as well as finiteness of correlation functions require 
that  \cite{FG, PSZ}
\begin{equation}
\gym^{2}=|h|^{2}\left(1-\frac{1}{N^{2}}|q-q^{-1}|^{2}\right) \ .
\label{vanishingbeta}
\end{equation}
This relation is expected to receive higher-loop corrections as well as corrections 
depending on higher powers of $1/N^2$.

\subsection{Coulomb branch}

The structure of the Coulomb branch of the $\beta$-deformed $\NeqFour$ theory was discussed in detail in 
\cite{Dorey:2003pp, Dorey:2004xm} and certain terms in the effective action for the light fields were 
evaluated through two loops in \cite{Kuzenko:2007tf}. As usual, the classical vacuum structure of the theory 
is determined by the $F$-  and $D$-term equations
\be
&&[\phi^{23},\phi^{31}]_\beta=0=[\phi^{31},\phi^{12}]_\beta=0=[\phi^{12},\phi^{23}]_\beta
\\
&&\epsilon_{ijk}[\phi^{ij},{\phi}^{k4}]=0 \ .
\nonumber
\ee
As it is well-known, for $\beta=0$ the solution to these equations is given by generic diagonal 
unitary matrices of unit determinant. At a generic point on the Coulomb branch the $SU(N)$ 
gauge group is broken to $U(1)^{N-1}$. 
A single scalar field with non-trivial vacuum expectation value -- {\it e.g.} 
$\phi^{23}={\rm diag}(\lambda_1,\dots,\lambda_N)$ -- is sufficient to break completely 
the gauge symmetry.  All fields except those valued in the Cartan subalgebra $U(1)^{N-1}$
become massive with masses given by:
\be
\label{neq4masses}
m^W_{ij}= |\lambda_i-\lambda_j|
~,~~
|m^\Psi_{ij}|= |\lambda_i-\lambda_j|
~,~~
|m^\phi_{ij}|= |\lambda_i-\lambda_j| \ .
\ee
Due to the extended supersymmetry these expressions do not receive corrections 
to any order in perturbation theory. 

For nonvanishing and generic\footnote{Interesting features emerge if 
$\beta/(2\pi)$ is rational.} value of $\beta$ the number of vacua is smaller; they are all 
inherited from the vacua of the undeformed theory.  The vacuum described above, in which 
only one scalar field has nontrivial vacuum expectation value, continues to exist. 
The masses of the $W$-bosons is unaffected by the deformation, while the masses of the 
charged fields become
\be
\label{masswdef}
m^W_{ij}= 
m^{\lambda^4}_{ij}= |\lambda_i-\lambda_j|
~~~~
m^{\phi^{23}}_{ij}= 
m^{\lambda^1}_{ij}= |\lambda_i-\lambda_j|
\\
m^{\phi^{13}}_{ij}=m^{\phi^{12}}_{ij}=m^{\lambda^{2}}_{ij}=m^{\lambda^{3}}_{ij}
= \frac{1}{\gym}\,h(\gym, N, q)\,|q\lambda_i-q^{-1}\lambda_j|
\label{massphidef}
\ee
Since the ${\cal N}=1$ supersymmetry algebra does not have a central charge, the 
relation between vacuum expectation values and masses of fields may receive quantum 
corrections. 

The IR divergences of scattering amplitudes of the undeformed theory may be 
regularized -- at least in the planar limit -- by evaluating them at a generic point 
on the Coulomb branch where the vacuum expectation value of scalar field(s) acts as 
a regulator. The nonrenormalization of eqs.~(\ref{neq4masses}) implies 
the extraction of the finite part of amplitudes through the subtraction of the known 
form of IR divergences  \cite{Akhoury:1978vq, Mueller:1979ih, Collins:1980ih, Sen:1981sd, 
Sterman:1986aj, Botts:1989nd, Catani:1989ne, Korchemsky:1988hd, Magnea:1990zb, 
Korchemsky:1993uz, Catani:1998bh, Sterman:2002qn, Collins:1989gx, Sen:1982bt, 
Aybat:2006mz} is straightforward.
This is no longer so in the presence of the deformation. Indeed, since (\ref{massphidef})
receive corrections at least through the condition that the theory is finite (eq.~(\ref{vanishingbeta}) at one- 
and two-loop orders)),
the universal IR divergences that should be subtracted will in fact depend on these corrected 
masses. It is moreover in principle possible that masses of the vector and chiral multiplets 
receive different finite renormalization. 
These new effects will modify the finite part of some $L$-loop amplitude by terms
proportional to lower-loop amplitudes and, for $\beta\in \IR$, are expected to appear 
at least at ${\cal O}(1/N)$ since conformal invariance requires 
$|h|^2=\gym^2$ \cite{Mauri:2005pa, Ananth:2006ac} to all orders in planar 
perturbation theory.

\subsection{The tree-level amplitudes of the $\beta$-deformed 
                    $\NeqFour$ sYM theory \label{trees}}

The interpretation of the $\beta$-deformed theory as a non-commutative deformation implies that
most of the tree-level amplitudes of the deformed theory are inherited from those of the 
$\NeqFour$ sYM theory. The presence of the double-trace terms in the component Lagrangian
signals however that there exist additional amplitudes, not present in the undeformed theory. 
To all-loop orders, the color decomposition of an $SU(N)$ gauge theory is \cite{Bern:1990ux}
\be
{\cal A}^\beta(k_1,\dots,k_n)&=&\sum_{S_n/Z_n}\Tr[T^{a_{\sigma(1)}}\dots T^{a_{\sigma(n)}}]
A^\beta_{n;1}(k_{\sigma(1)},\dots, k_{\sigma(n)})
\\
&
+&\!\!\!\!\!\!\sum_{S_n/Z_{n_1}\times Z_{n-n_1}}\!\!\!\!
\Tr[T^{a_{\sigma(1)}}\dots T^{a_{\sigma(n_1)}}]\Tr[T^{a_{\sigma(n_1+1)}}\dots T^{a_{\sigma(n)}}]
A^\beta_{n;2}(k_{\sigma(1)},\dots, k_{\sigma(n)})
\nonumber\\
&+&\text{terms with}~m\ge 3~\text{traces} +{\cal O}(1/N)\ .
\nonumber
\ee
In the following we will {sometimes} use the shorthand notation
$\Tr[T^{a_1}\dots T^{a_n}]\equiv \Tr_{1\dots n}$.
%
%
The  tree-level single-trace terms to leading order in $1/N$ are inherited from the
$\NeqFour$ theory\footnote{The same is true for the leading terms in the $1/N$ expansion 
of the single-trace partial amplitudes to all orders in perturbation theory \cite{Khoze:2005nd}.}:
\be
A^{\beta,(0)}_{n;1}(k_{\sigma(1)},\dots, k_{\sigma(n)})
=e^{i\varphi(\q_{\sigma(1)},\dots,\q_{\sigma(n)})}
A^{\NeqFour,(0)}_{n;1}(k_{\sigma(1)},\dots, k_{\sigma(n)}) 
\ee
where the phase $\varphi$ is defined in eq.~(\ref{phase_general}):
\be
\varphi(\q_{1},\dots,\q_{n})=\sum_{i=1}^n\sum_{j=i+1}^n
\beta_{ab}\q^a_{i}\q^b_{j}=\sum_{i=1}^n\sum_{j=i+1}^n
\q_{i}\wedge \q_{j}~~.
\label{fullphase1tr}
\ee
Clearly, the phase $\varphi$ depends on the color ordering. The antisymmetry of $\beta$ 
ensures that
\be
\varphi(\q_1,\dots,\q_n)=-\varphi(\q_n,\dots,\q_1) \ ;
\ee
this property is crucial for the finiteness of the theory. Tree-level $1/N^2$ terms in the
single-trace sector appears because the coefficient of the superpotential differs from the gauge 
coupling at this order (\ref{vanishingbeta}).

While the noncommutative interpretation of the $\beta$-deformation is transparent in
the trace-based color decomposition, a color decomposition based on the structure 
constants is possible. Such a decomposition also makes contact with color/kinematics
duality \cite{Bern:2008qj} and allows us to examine its fate once $\beta\ne 0$. From this perspective, 
the main consequence of the deformation is the appearance of the symmetric structure constants 
$d_{abc}$ in the coupling of scalar fields, either among themselves or with fermions. The 
structure constant in the three-point vertices with all changed fields are then replaced with
\be
\label{fbeta}
{\rm f}^{abc}_\varphi = \Tr[T^a[T^b,T^c]_\beta] &=& e^{i\varphi(a,b,c)}\Tr[T^aT^bT^c]-
e^{i\varphi(a,c,b)}\Tr[T^aT^cT^b] \cr
&=& \cos(\varphi(a,b,c))f^{abc}+i\sin(\varphi(a,b,c))d^{abc}
\ee
where the phase $\varphi$ is determined by the fields attached to the vertex. 
These modified structure constants are antisymmetric if, when interchanging 
two color indices, one also interchanges the R-charges of the corresponding fields; this is 
possible because the R-charge is conserved at each three-point vertex. 
%
%
It is not difficult to see that multi-trace tree-level amplitudes contain an even number of such modified 
structure constants.

\subsection{Examples: four-point amplitudes at tree-level \label{examples}}

As we will discuss in detail in \S~\ref{susy_Ward}, four-point amplitudes in 
the $\beta$-deformed $\NeqFour$ theory can be classified following the number 
and position of fields in the vector multiplet. At tree-level the situation is more constrained,
as most amplitudes are closely related to those of the undeformed theory and, as such, 
superficially enjoy all the constraints imposed by $\NeqFour$ supersymmetry. 

By considering the R-charges of fields it is not difficult to see that only single-trace
amplitudes with at most one field in the vector multiplet,
\be
\label{modamps}
{\cal A}_{4}^{(0)}(1g^+,2\phi^{23},3\psi^{134},4\psi^{124})
~~
\text{and}
~~
{\cal A}_{4}^{(0)}(1\phi^{23},2\phi^{14},3\phi^{13},4\phi^{24})  \ ,
\ee
receive $\beta$-dependent corrections. For all other field configurations 
the antisymmetry of the phase (\ref{fullphase1tr}) implies that any potential 
$\beta$-dependent phase factor is absent; in general at least three different nontrivial 
charge vectors are necessary for a single-trace tree-level amplitude to be affected by 
the deformation. At higher loops differences appear for other external field configurations. 

All color-ordered amplitudes included in the first amplitude in eq.~(\ref{modamps}) are modified either by 
a factor of $q=\exp(i\beta)$ or its inverse -- for example
\begin{eqnarray}
\nonumber
{A}_{4;1}^{\beta,(0)}(1234)&=&\h(\gym, N,q)\,  q\, A_{4;1}^{\NeqFour,(0)}(1234)
\\
{A}_{4;1}^{\beta,(0)}(1243)&=&\h(\gym, N,q)\,  q^{-1}\,A_{4;1}^{\NeqFour,(0)}(1243) \ ,
\label{treegsff}
\end{eqnarray}
where $\h$ is defined as $\h = h/\gym$. For this field configuration there is no double-trace 
tree-level amplitude.
%
%

For the second amplitude in (\ref{modamps})  all single-trace color-ordered 
amplitudes are the same as in the $\NeqFour$ theory except for
\begin{eqnarray}
{A}_{4;1}^{\beta,(0)}(1324)&=& \h(\gym, N,q)^{2}\,q^{2}{A}_{4;1}^{\NeqFour,(0)}(1324)\\
{A}_{4;1}^{\beta,(0)}(1342)&=& {A}_{4;1}^{\NeqFour,(0)}(1342)
+(1-\h(\gym, N,q)^{2}){A}_{4;1}^{\NeqFour,(0)}(1324) \ .
\end{eqnarray}
In the first case the complete amplitude comes from a single $\beta$-dependent 
four-scalar vertex. In the second, the amplitude receives two different contributions: from a 
$\beta$-independent 4-scalar vertex and from two three-point gluon scalar vertices, 
the former being equal to $-\h(\gym, N,q)^2 {A}_{4;1}^{\NeqFour,(0)}(1324)$.
This amplitude also contains a double-trace component:
\be
{A}_{4;2}^{\beta,(0)}(13;24)&=&-\frac{1}{N}\h(\gym, N,q)^{2}(q-q^{-1})^{2} \ .
\label{2trssss}
\ee
This term is generated by the double-trace Lagrangian ${\cal L}_2$ in eq.~(\ref{def_lag}). 
Supersymmetry requires that there exist similar 
two-trace amplitudes with two fermions and two scalars as well as   
two-trace amplitudes with four external fermions. They may be generated 
from (\ref{2trssss}) though supersymmetry Ward identities or by direct evaluation 
starting from the Lagrangian (\ref{def_lag}) and focusing on the terms 
containing the symmetric structure constants.
Higher-point multi-trace amplitudes can be found though {\it e.g.} color-dressed on-shell 
recursion relations or color-dressed MHV diagrams in a trace basis. 

There exist, of course, other presentations of color-dressed amplitudes of this theory.
In particular, to examine the fate of the color/kinematics duality for the 
$\beta$-deformed theory it is useful to express the color factors in terms of 
structure constants. Both the modified structure constants (\ref{fbeta}) as well as the 
standard antisymmetric structure constants are necessary, the former coming from the 
interaction terms depending solely on the chiral multiplets and the latter from the vector 
multiplet interactions.

Using the modified structure constants (\ref{fbeta}) it is  not difficult to see that 
the representatives (\ref{modamps}) of the two classes of amplitudes which 
receive $\beta$-dependent deformations
may be written as
\be
\label{modamp1}
{\cal A}_{4}^{\beta,(0)}(1g^+,2\phi^{23},3\psi^{134},4\psi^{124})&=&
\h\frac{n_{12}}{s_{12}}f^{12a}f_\beta^{34}{}_{a}
+
\h\frac{n_{23}}{s_{23}}f_\beta^{23a}f^{14}{}_{a}
+
\h\frac{n_{13}}{s_{13}}f^{31a}f_\beta^{24}{}_{a}
\\[3pt]
{\cal A}_{4}^{\beta,(0)}(1\phi^{23},2\phi^{14},3\phi^{13},4\phi^{24})&=&
\frac{n'_{12}}{s_{12}}f^{12a}f^{34}{}_{a}
+
\frac{n'_{23}}{s_{23}}f^{23a}f^{14}{}_{a}
+
|\h|^2\frac{n'_{13}}{s_{13}}f_{-\beta}^{31a}f_\beta^{24}{}_{a} \ ,
\label{modamp2}
\ee
where $\h$ stands for $\h(\gym, N,q)$ and $f_\beta$ is ${\rm f}_\varphi$ for $\varphi=\beta$.
The third term in the second equation above includes the double-trace 
partial amplitude in (\ref{2trssss}).  
In eqs.~(\ref{modamp1}) and (\ref{modamp2}) the various numerator factors $n_{ij}$ have a 
Feynman diagram interpretation, being determined by the three- and four-point vertices from 
the Lagrangian (\ref{def_lag}) according to their color factors.
In general, the $\beta$-modified color factors do not obey the Jacobi identity; in the special 
cases when they do, such as  the eq.~(\ref{modamp1}), they may be modified as in the undeformed 
theory by adding to the amplitude a general function multiplied by the sum of all color factors.
In such cases one may choose the numerator factors $n_{ij}$ to be the same as in the undeformed 
theory and consequently to obey the Jacobi-like relation
\be
n_{12}+n_{23}+n_{13}=0 \ .
\ee
%
It is perhaps worth mentioning that the appearance of the symmetric invariants $d_{abc}$ 
prevents use of the multi-peripheral color decomposition of \cite{DelDuca:1999rs}, which uses 
Jacobi identities to cast the color factors in a specific form.

\section{Loop calculations in $\beta$-deformed $\NeqFour$ sYM theory: \\
general strategy and all-order properties \label{general_loops}} 

We will construct color-dressed loop amplitudes in the $\beta$-deformed $\NeqFour$ sYM 
theory using the generalized unitarity-based method \cite{Bern:1994zx, Bern:1994cg, 
Bern:1997sc, Britto:2004nc, Buchbinder:2005wp}. 
Its key feature, that it uses tree-level amplitudes as input for higher-loop calculations, 
makes it particularly useful in our case and reduces calculations to dressing the contributions
to generalized unitarity cuts with the appropriate $\beta$-dependent factors.
\footnote{For example, one may easily see that, in all planar cuts, the complete 
$\beta$-dependence on internal lines cancels out and these cuts are the same as 
in the undeformed theory up to an overall phase independent of the cut lines. 
This recovers the results of \cite{Khoze:2005nd} for real $\beta$. The argument fails for 
complex $\beta$ since in that case $h(\gym, N, q) \ne \gym$ even to leading order in $1/N$; 
it is instead given by $\gym^2=|h|^2(|q|^2+|q^{-1}|^2)$ \cite{Khoze:2005nd}.}

\subsection{Generalities}

In general, subleading color amplitudes depend nontrivially on the deformation parameter,
feature that may be already seen at the level of generalized cuts. 
An important ingredient in their evaluation is the sum over intermediate states -- the 
supersum. For $\NeqFour$ sYM theory efficient methods for their evaluation -- both 
graphically and algebraically -- have been described in \cite{Bern:2009xq}.\footnote{
For pure sYM theories with reduced supersymmetry these methods have been organized 
in \cite{Elvang:2011fx} in terms of a reduced-supersymmetry superspace. }
The graphical method together with knowledge of the general structure of supersums identified 
a set of rules which allow the construction of the supersum in terms of only the purely gluonic 
intermediate states and the intermediate states containing gluons and two gauginos. 
%
%
In our case, the graphical method, which tracks the R-charge flow across generalized cuts, seems the
ideal approach\footnote{It may nevertheless be possible that some modification of the 
algebraic approach could be devised.} as it allows us to dress each contribution to the supersum with the 
appropriate phase factors and to incorporate the changes to the coefficient of the superpotential 
(\ref{vanishingbeta}). Indeed, one may split the  sum over intermediate states into the contributions 
of the various $\NeqOne$ multiplets and subsequently include the contribution of the 
deformation.

Let us consider here a simple example -- the $s_{12}$ cut of 
${\cal A}^{\beta,(1)}_4(1g^+,2g^-, 3\phi^{23},4\phi^{14})$ which will be necessary 
for the construction of this amplitude in the next section.
The general structure of this cut 
is\footnote{As it is well-known \cite{Bern:1994cg}, one-loop 
amplitudes in supersymmetric gauge  theories are determined, through $O(\epsilon)$,  by their 
four-dimensional cuts. 
}
\be
{\cal C}=\rho^2\;\frac{\langle l_a 2 \rangle^2\langle l_b 2\rangle^2}
                           {\langle 1 2\rangle\langle 2 l_a\rangle\langle l_a l_b \rangle\langle l_b 1\rangle}
                         \frac{\langle l_b 4 \rangle \langle l_b 3 \rangle\langle l_a 4 \rangle \langle l_a 3 \rangle}
                           {\langle 3 l_a\rangle\langle l_a 4\rangle\langle4 l_b\rangle\langle l_b 3\rangle}
\ee
where $1,2,3,4$ stand for the appropriate spinors of external momenta 
and $\rho$ captures the sum over intermediate states. For $\NeqFour$ sYM theory $\rho$ is
\be
\rho^2_{\NeqFour}=A^2 B C-(2 A B+2AC) +(4+\frac{B}{C}+\frac{C}{B})-(\frac{2}{A B}+\frac{2}{AC}) + \frac{1}{A^2 BC}
\label{ss}
\ee
where the five terms correspond to two-gluon, fermion, two-scalar, two-fermion and two-gluon intermediate 
states, respectively, and 
\be
A=\frac{\langle l_a 2\rangle}{\langle l_b 2\rangle}
~~~~
B=\frac{\langle l_b 3\rangle}{\langle l_a 3 \rangle}
~~~~
C=\frac{\langle l_b 4\rangle}{\langle l_a 4 \rangle} ~~.
\ee
%
%
One half of the first term in the contribution of the two-fermion intermediate state 
is due to the fermion in the vector multiplet  and thus cannot acquire 
$\beta$-dependence; the second half comes from fermions in the same 
chiral multiplet as the external scalars and, consequently, it cannot have any 
$\beta$-dependence either.
One may similarly 
identify the two non-constant terms in the third parenthesis as coming from an 
intermediate state with the same R-charge as the external scalars. Thus, the modified 
supersum is
\be
\rho^2_\beta\!\!\!&=&\!\!\!A^2BC-(2AB+2\cos(2\beta)  AC)
+(4\cos(2\beta)+\frac{B}{C}+\frac{C}{B})
-(\frac{2}{AB}+\frac{2\cos(2\beta) }{AC} )
+ \frac{1}{A^2BC}\cr
&=&\!\!\!\rho^2_{{\cal N}=4}+4\left(\sqrt{AC}-\frac{1}{\sqrt{AC}}\right)^2\sin^2\beta
\label{sseg}
\ee
In general, one may either follow the same strategy as here or simply consider all index 
diagrams \cite{Bern:2009xq} and for each of them construct the phase factor of the 
amplitude factors while taking into account the color ordering.

The structure of a supersum reflects the supersymmetry preserved by the corresponding
generalized cut; for example, in the $\NeqFour$ theory they are perfect fourth powers, signaling the 
fact that none of the supercharges is broken. In the example in eq.~(\ref{sseg}) we notice that 
the supersum is a perfect square; this suggests that, in some sense, this cut preserves 
${\cal N}=2$ supersymmetry. While this feature appears for other cuts of one-loop amplitudes,
it is not a universal phenomenon. Its interpretation and 
consequences are also not completely clear; a possibility is that, 
while some cuts of four-point amplitudes preserve a larger amount of supersymmetry, only four 
supercharges are common to all of them.

The antisymmetry of the Moyal product (\ref{Moyal}) introduces a certain correlation 
between the order of fields and the  $\beta$-dependence. Because of this we will need to 
identify the cuts of color-stripped amplitudes by extracting them from color-dressed cuts.
This may be done either in the trace basis or in terms of the modified 
structure constants (\ref{fbeta}). We will organize the results in the trace basis and express 
them at the end in terms of structure constants. An advantage of this approach is that 
we avoid the potential proliferation of color factors due to possible orderings of undeformed 
and deformed structure constants $f$ and $f_\beta$, respectively.  
Since multi-trace amplitudes already appear at tree level, the loop level color decomposition of 
amplitudes is essentially the same as the one discussed in \S~\ref{trees}.\footnote{As usual, 
the structure of tree-level multi-trace terms implies certain constraints on the IR divergences 
of loop amplitudes in the deformed theory \cite{Akhoury:1978vq}-\cite{Aybat:2006mz}.}

In the next sections we will determine all one-loop four-point amplitudes as well as certain 
two-loop four-point amplitudes and analyze their properties. To isolate the effects of the 
deformation it is convenient to separate,  at each loop order, the  $\NeqFour$ part of 
amplitudes:
\be
\label{separate}
{\cal A}_n^{\beta,(L)}={\cal A}_n^{\NeqFour,(L)}+{\cal A}_n^{\text{extra},(L)} \ ;
\ee
the second term, ${\cal A}_4^{\text{extra},(L)}$, vanishes as some power of $\sin^2\beta$.
The presence of the double-trace tree-level amplitudes as well as the expression 
(\ref{vanishingbeta}) for the coefficient of the superpotential are crucial for the UV 
finiteness of the theory in four dimensions.

\subsection{The conformal invariance condition}

While the arguments of Leigh and Strassler \cite{Leigh:1995ep} guarantee that there exists
a normalization of the superpotential which guarantees conformal invariance, the expression
of this coefficient in terms of $\beta$, $N$ and the gauge coupling is not known {\it a priori}. 
As reviewed in \S~\ref{review}, the coefficient $h$ solving the eq.~(\ref{vanishingbeta}) 
required by a vanishing one-loop $\beta$-function also implies that all UV divergences cancel 
at two loops as well. As we will see here, a simple color-based argument implies that this is only 
an accident and the coefficient $h$ receives corrections at the next order, {\it i.e.} at three loops.

It is not difficult to devise an argument which recovers, without any explicit calculations,
the one- and two-loop constraint (\ref{vanishingbeta}). To this end, let us consider supergraphs
correcting the scalar field propagator which do not contain any internal vector multiplets. 
Any divergence appearing in these graphs -- which are essentially the graphs of a particular 
three-flavor Wess-Zumino model -- must be cancelled by graphs containing vector multiplet 
lines, both in the presence and in the absence of the deformation. Using this constraint we can
reconstruct the divergence of graphs of the latter type from the divergence of graphs of the 
former type.

\begin{figure}
\centering
{\includegraphics[height=24mm]{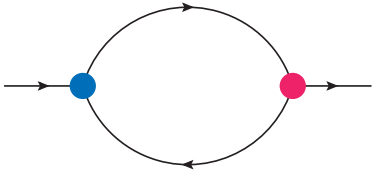}} 
\caption{One-loop correction to the scalar propagator arising from the interactions 
of chiral multiplets.}
\label{1loopprop}
\end{figure}

At one-loop level there exists a single purely scalar superfield propagator correction, shown 
in fig.~\ref{1loopprop}; it contains one chiral and one anti-chiral vertex, each of which is 
proportional to the modified structure constant $f_\beta$ found in eq.~(\ref{fbeta}). 
Evaluating the sum over the color indices it is not difficult to find that
\be
{\rm fig.}~\ref{1loopprop}\propto |h|^2\left(1-\frac{1}{N^2}|q-q^{-1}|^2\right) \ .
\ee
Finiteness of the 1-loop correction to the chiral superfield propagator requires that, after the 
contribution of the internal vector multiplet is added, the UV divergence of the scalar two-point
function is proportional to 
\be
\langle \Phi^i(p){\bar \Phi}_j(-p)\rangle\big|_{\rm 1-loop} \propto \frac{1}{\epsilon}\,\delta^i_j \,p^2 \,
\left[\gym^2 - |h|^2\left(1-\frac{1}{N^2}|q-q^{-1}|^2\right)\right] \ ;
\ee
requiring that the divergence vanishes implies immediately that eq.~(\ref{vanishingbeta}) must be satisfied.

\begin{figure}
  \centering
  \subfigure[]{%
    \includegraphics[height=22mm]{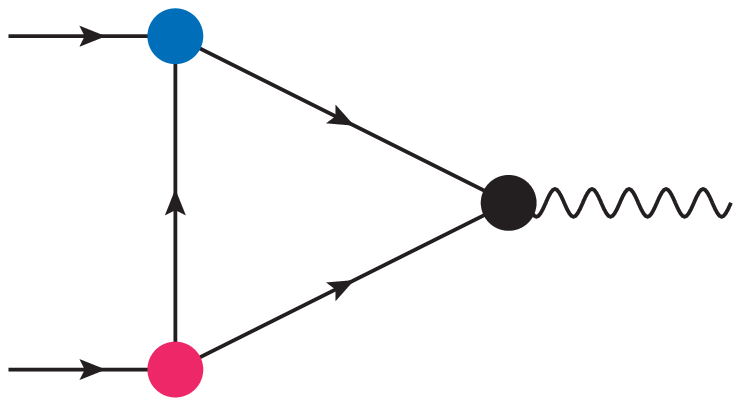}
    \label{vert1}
  }
  ~~~~~~~
  \subfigure[]{%
    \includegraphics[height=22mm]{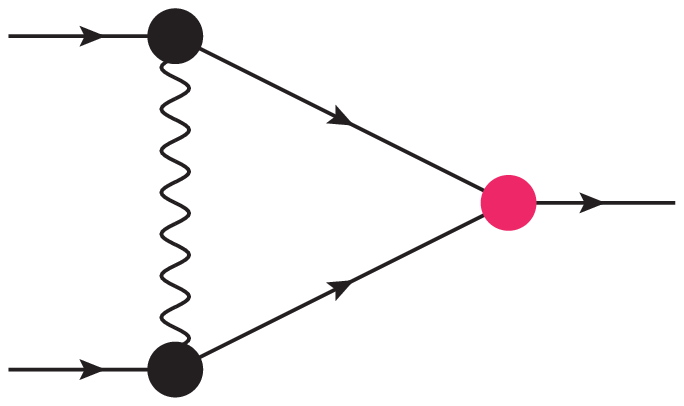}
    \label{vert2}
  }
  \caption{One-loop correction to three-point vertices.}
  \label{1loopvertex}
\end{figure}

This argument may be extended to the two-loop correction to the propagator of the chiral 
superfields. To this end we notice that $U(1)$ charge conservation forbids any contributions
containing only scalar interactions. We are therefore to identity the diagrams containing the 
smallest number of internal vector multiplet lines; it is straightforward to see that one such line 
is sufficient.
The second important observation is that any such two-loop propagator correction necessarily 
contains a triangle subintegral. The two possible subintegrals are shown in fig.~\ref{1loopvertex}; 
their corresponding color structures are 
\be
{\rm fig.}~\ref{vert1}\propto \,\gym\,|h|^2\,\left(1-\frac{1}{N^2}|q-q^{-1}|^2\right)  f^{abc}
\quad , \quad
{\rm fig.}~\ref{vert2}\propto \,\gym^2 \,h \,f_{\beta}^{abc} \ .
\ee 
This implies that, from a color space perspective, two-loop propagator corrections reduce to a 
one-loop analysis. Moreover, the vertex correction in fig.~\ref{vert1} generates an additional
$q$-dependent factor when inserted into a propagator correction. We therefore conclude that 
the divergent part of the two-loop correction to the scalar superfield propagator due to
diagrams with the smallest number of vector multiplets is proportional to 
\be
\frac{1}{\epsilon}\,\delta^i_j \, p^2\, \gym^2 |h|^2\left(1-\frac{1}{N^2}|q-q^{-1}|^2\right) \ .
\ee
Requiring that this divergence is cancelled at $q=1$ by diagrams with further vector multiplet 
lines implies that 
\be
\langle \Phi^i(p){\bar \Phi}_j(-p)\rangle\big|_{\rm 2-loop}\propto \frac{1}{\epsilon}\delta^i_j 
\,p^2 \,\gym^2\, \left[\gym^2-|h|^2\left(1-\frac{1}{N^2}|q-q^{-1}|^2\right)\right] \ .
\ee
Thus, the condition of one-loop finiteness also implies finiteness at two loops.

\begin{figure}
\centering
{\includegraphics[height=24mm]{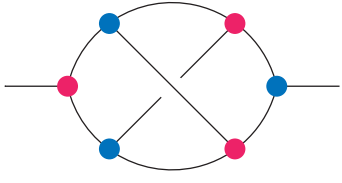}} 
\caption{Three-loop correction to the scalar propagator arising from the interactions 
of chiral multiplets and containing only box sub-integrals. This color space contribution 
is not reducible to lower-loop structures.}
\label{3loopprop}
\end{figure}

The same type of arguments imply that, at three loop order, there exists a color structure 
which is different from those that appear at one and two loops.  This is a consequence of 
the fact that at this loop order there exists a Feynman diagram, shown in fig.~\ref{3loopprop},
which does not contain any triangle subintegrals and hence cannot be reduced to lower-loop 
structures. It is not difficult to find that 
\be
{\rm fig.}~\ref{3loopprop}\propto 
|h|^6 \left( N-\frac{4}{N}\right)|q-q^{-1}|^4\left(\left(q^2+4+q^{-2}\right)+\frac{5}{N^2}|q-q^{-1}|^2\right) \ .
\ee
Clearly, such a divergence cannot be cancelled unless the coefficient $h$ receives further 
corrections compared to its one- and two-loop expression (\ref{vanishingbeta}). Apart from the 
three-loop propagator corrections, to determine the change to the conformal invariance 
condition (\ref{vanishingbeta}) the precise form of the two-loop divergence is also necessary.

\subsection{Supersymmetric relations \label{susy_Ward}}

Supersymmetry imposes strong constraints on the structure of amplitudes in the 
$\beta$-deformed theory; these constraints are, to some extent, common to all $\NeqOne$ 
supersymmetric theories. Since we will be mostly interested in four-point amplitudes, let us 
review the consequences of supersymmetry Ward identities have on them.

In the $\NeqFour$ theory all four-point amplitudes fit inside a single superamplitude, 
{\it i.e.} they are all related to each other by $\NeqFour$ supersymmetry transformations. 
Color-ordered four-point superamplitude is given by \cite{Bern:1996ja, Alday:2007hr}
\be
{\bf A}_4^{\NeqFour} = {\bf A}_4^{\NeqFour,(0)}M_4(s, t) \ ,
\ee
where ${\bf A}_4^{\NeqFour,(0)}$ is the tree-level color-ordered superamplitude.

In the presence of the deformation the various helicity states can be organized in 
four $\NeqOne$ superfields and their conjugates:
\be
&&
V=g^+ + \eta\psi^4
\qquad\quad
\Phi^{ij}=\phi^{ij} + \eta\, \psi^{ij4}~~~~~i,j=1,2,3 
\\
&&
V^\dagger=g^- + {\bar \eta}\, \psi^{ijk}
\qquad
\Phi^\dagger_{ij}={\bar \phi}_{ij} + {\bar \eta}\psi_{ij4}= \epsilon_{ijk}(\phi^{k4}+{\bar\eta}\psi^k) \ .
~~~~i\ne j\ne k
\label{superfields}
\ee
The superfield $V$ containing the gauge fields and the 
gauginos is denoted by $\Phi$ in \cite{Elvang:2011fx}; we have labeled the superfields 
containing the scalar fields and their partners with the same labels as those of the scalar 
fields in the $\NeqFour$  theory. To preserve the similarity with the $\NeqFour$ theory
one may additionally Fourier-transform ${\bar \eta}$. 

The $\NeqOne$ supersymmetry Ward identities \cite{SWI} are much less restrictive than their  
$\NeqFour$ counterparts. For example, since no sequence of transformations maps a 
positive helicity gluon into a negative helicity one, color-ordered amplitudes related by interchanging 
helicities of gluons -- {\it e.g.} $A(1g^+2g^+3g^-4g^-)$ and $A(1g^+2g^-3g^+4g^-)$ --
are not necessarily related beyond tree level.\footnote{Similarly to pure YM theory, 
at tree-level the $\beta$-deformed $\NeqFour$ theory is, effectively, maximally 
supersymmetric.} Consequently, up to CPT conjugation, the independent four-point amplitudes are
\be
&&
VVV^\dagger V^\dagger 
~,~
VV^\dagger \Phi\Phi^\dagger 
~,~
V\Phi \Phi'\Phi'' 
~,~ 
V\Phi^\dagger \Phi'^\dagger\Phi''^\dagger 
~,~
\Phi\Phi^\dagger \Phi\Phi^\dagger 
~,~
\Phi\Phi^\dagger \Phi'^\dagger\Phi' ~ ,
\label{fieldconfig}
\ee
together with all their noncyclic permutations. Here $\Phi$, $\Phi'$ and $\Phi''$ stand 
for different chiral multiplets (\ref{superfields}).
Conservation of the three $U(1)$ charges introduces the same selection rules as the $SU(4)$
R-symmetry of the undeformed theory; in particular, it requires that amplitudes with three 
vector multiplets and one chiral multiplet vanish identically. For this reason field 
configurations with nonvanishing charges are not listed in eq.~(\ref{fieldconfig}). 

At the expense of introducing spurious poles in external momentum invariants, one
may always extract some ratio of spinor products which accounts for the phase weight 
determined by the helicities of the external fields. Through two-loop level, it appears however 
that the following relations are naturally satisfied:
\be
VVVV: &&
\mathcal{A}^{\text{extra}}_4(V_1V_2^\dagger V_3V_4^\dagger)\propto 
\mathcal{A}^{\NeqFour,(0)}_4(V_1V_2^\dagger V_3V_4^\dagger)
\nonumber\\[3pt]
VV\Phi\,{\Phi}^\dagger: &&
\mathcal{A}^{\text{extra}}_4(V_1V_2^\dagger \Phi_3\Phi_4^\dagger)\propto
\frac{\langle 23\rangle^2}{\langle 13\rangle^2}
\nonumber\\[3pt]
V\Phi\,\Phi\,\Phi: &&
\mathcal{A}^{\text{extra}}_4(V_1\Phi_2 \Phi_3\Phi_4)\propto
\mathcal{A}^{\beta,(0)}_4(V_1\Phi_2 \Phi_3\Phi_4)
\nonumber\\[3pt]
\Phi\,\Phi^\dagger\Phi\,\Phi^\dagger: &&
\mathcal{A}^{\text{extra}}_4(\Phi_1\Phi^\dagger_2 \Phi_3\Phi^\dagger_4)\propto 1
\nonumber\\[3pt]
\Phi\Phi^\dagger\Phi'\Phi'^\dagger:  &&
\mathcal{A}^{\text{extra}}_4(\Phi_1\Phi^\dagger_2 \Phi'_3\Phi'^\dagger_4)\propto 1
\nonumber
\ee
with other field configurations being obtained by simple relabeling following the noncyclic 
permutations of external fields.

\subsection{Non-planar inheritance 
\label{allorders}}

While at higher orders in the $1/N$ expansion amplitudes in the deformed theory differ 
substantially from those of the undeformed one, the leading term in the multi-color expansion
of certain multi-trace amplitudes with special configurations of external R-charges is 
nevertheless inherited from the undeformed theory.

\begin{figure}
\centering
{\includegraphics[height=40mm]{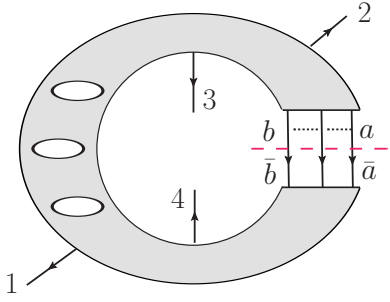}} 
\caption{Generalized cut showing that, if $q_1+q_2=-q_3-q_4=0$, the leading $1/N$ term
in the $\Tr_{12}\Tr_{34}$ double-trace amplitude in the $\beta$-deformed theory  
is proportional to the corresponding $\NeqFour$ amplitude.}
\label{fig:2trace_nobeta}
\end{figure}

Let us consider an $L$-loop single-trace planar scattering amplitude with $(2m+n)$ external 
lines and let us assume that the R-charges of the $(2m)$ external lines are such that this 
amplitude contributes to an $m$-particle cut of an $m$-loop amplitude. We will focus on the leading 
double-trace terms captured by such a generalized cut, which is illustrated in 
fig.~\ref{fig:2trace_nobeta} for $n=4$. 
We will see that, for a suitable choice of R-charges for the 
external lines, all $\beta$-dependence drops out of this cut, implying that the particular
double-trace structure captured by it is unaffected by the deformation. 

Let us therefore analyze the phase of the relevant tree-level graph and cast it into a form suitable for
the evaluation of this cut, which singles out $\Tr[T^{a_1}\dots T^{a_p}]
\Tr[T^{a_{p+1}}\dots T^{a_n}]$. Denoting the cut legs by $a_i$ and ${\bar a}_i$ with 
$i=1,\dots,m$ on the two sides of the cut (suggesting the fact, to be used shortly, that the 
R-charges of sewn legs are equal in magnitude and opposite in sign), the phase is 
\be
&&\varphi(1,\dots,p, a_1,\dots,a_m, p+1,\dots, n, {\bar a}_{m},\dots, {\bar a}_1) 
\nonumber\\
&=& 
\sum_{i=1}^p \left(q_i \wedge \sum_{j=i+1}^p q_j+q_i \wedge\sum_{j=1}^m  q_{a_j}+
q_i \wedge\sum_{j=p+1}^n q_j + q_i \wedge\sum_{j=m}^1 q_{{\bar a}_j}\right)
\cr
&+&
\sum_{i=1}^m \left(q_{a_i} \wedge\sum_{j=i+1}^m  q_{a_j}+
q_{a_i} \wedge\sum_{j=p+1}^n  q_j+q_{a_i} \wedge\sum_{j=m}^1 q_{{\bar a}_j}\right)
\cr
&+& 
\sum_{i=p+1}^n \left(q_i \wedge \sum_{j=i}^n  q_j
+q_i \wedge \sum_{j=m}^1 q_{{\bar a}_j}\right)
\cr
&+& 
\sum_{i=m}^1 \left(\sum_{j=i-1}^1 q_{{\bar a}_i} \wedge q_{{\bar a}_j}\right) \ .
\label{fullphase}
\ee
Charge conservation removes all dependence of the charge of the cut legs from the first line
above; for the same reason, the terms on the second line may also be written as
\be
\sum_{i=1}^m \left(-q_{a_i}\wedge\sum_{j=1}^p q_j - q_{a_i}\wedge \sum_{j=1}^{i-1}q_{a_j}\right) \ ;
\label{line2}
\ee
On the third line, the dependence on the charges of cut legs enters multiplied 
by $\sum_{i=p+1}^n q_i$. Finally, the sum on the fourth line may be reorganized as
\be
&&\sum_{i=m}^1 \sum_{j=i-1}^1 q_{{\bar a}_i} \wedge q_{{\bar a}_j} 
=\sum_{j=1}^m \sum_{i=m}^{j+1} q_{{\bar a}_i} \wedge q_{{\bar a}_j}  
\\
&=& 
+\sum_{j=1}^m\sum_{i=1}^p q_{i} \wedge q_{{a}_j}
+\sum_{j=1}^m\sum_{i=1}^m q_{a_i} \wedge q_{{a}_j}
+\sum_{j=1}^m\sum_{i=p+1}^n q_{i} \wedge q_{{ a}_j}
-\sum_{j=1}^m \sum_{i=1}^{j-1} q_{{a}_i} \wedge q_{{a}_j}  \ ,
\label{line4}
\ee
where we also used the relation $q_{\bar a}=-q_a$. The first term on the second line combines with the first term on the right-hand side of (\ref{line2}). 
The second term on the second line vanishes, since the two charge vectors are equal. 
The third term on the second line cancels the second term on the third line 
of (\ref{fullphase}); last, the fourth term on the second line in (\ref{line4}) cancels the second 
term in (\ref{line2}). Combining everything, it is not difficult to see that
\be
\varphi(1,\dots,p, a_1,\dots,a_m, p+1,\dots, n, {\bar a}_{m},\dots, {\bar a}_1)
=\sum_{i=p+1}^n \sum_{j=i}^n q_i \wedge q_j 
+ \sum_{i=p+1}^n q_i \wedge \sum_{j=1}^m q_{a_j} \ .
\ee
Thus, if $\sum_{i=p+1}^mq_i = - \sum_{i=1}^p q_i = 0$, the $\beta$-dependent phase 
of the generalized cut contributing to the double-trace structure  $\Tr[T^{a_1}\dots T^{a_p}]
\Tr[T^{a_{p+1}}\dots T^{a_n}]$ does not depend on the charges of the cut lines; consequently,
this cut is the same as in the undeformed theory.

We may give a pictorial interpretation to this result. In color space, all Feynman 
diagrams contributing to the cut analyzed above have the topology of a cylinder, with 
legs $1,\dots,p$ attached to one boundary and legs $p+1,\dots,n$ attached to the other. 
If the total charge flowing in through one boundary, $\sum_{i=p+1}^mq_i = - \sum_{i=1}^p q_i$,
vanishes, then the corresponding double-trace amplitude is not deformed apart from a 
potential overall phase. Examples include 
$\Tr_{12}\Tr_{34}$
in $\mathcal{A}_{4}^{\beta}(1g^{+},2g^{-},3\phi^{ij},4\phi^{k4})$ (with $i\ne j\ne k=1,2,3$), 
%
$\Tr_{12}\Tr_{34}$ and $\Tr_{14}\Tr_{23}$
in
$\mathcal{A}_{4}^{\beta}(1\phi^{23},2\phi^{14},3\phi^{23},4\phi^{14})$
and 
%
$\Tr_{12}\Tr_{34}$
in
$\mathcal{A}_{4}^{\beta}(1\phi^{23},2\phi^{14},3\phi^{13},4\phi^{24})$.
This is a direct counterpart of a similar result proven in \cite{Chepelev:1999tt} for space-time 
non-commutative field theories. The explicit calculations in the following sections support the
conclusions reached here.

It is not difficult to generalize this criterion to all multi-trace amplitudes; their color-space 
graphs have the topology of a sphere with as many punctures as trace factors. If the R-charge 
flow through all punctures vanishes, then the  corresponding amplitude is the same as in the 
$\NeqFour$ theory even when $\beta\ne 0$. For example, all ${\cal O}(1/N)$ multi-trace 
gluon amplitudes are the same as in the undeformed theory; beginning at three-loop order, they
are more convergent in the UV than the single-trace leading color terms \cite{Bern:2010tq}.

\section{$\beta$-deformed $\NeqFour$ sYM theory at one-loop \label{oneloop}}

\begin{figure}
  \centering
  \subfigure[${\cal I}_{1,2,3,4}$]{%
    \includegraphics[height=22mm]{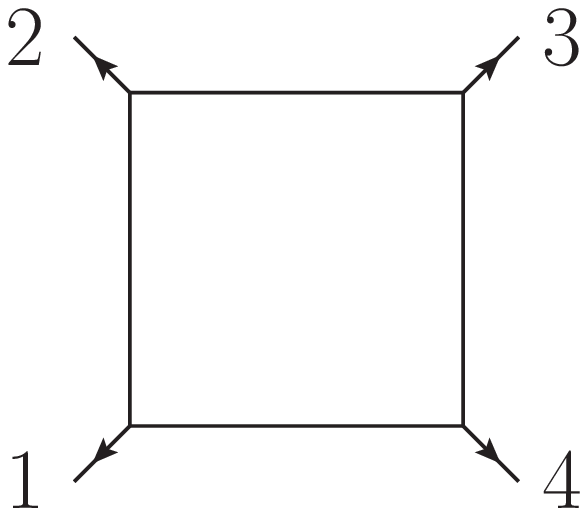}
    \label{fig:1}
  }
  ~~~~
  \subfigure[${\cal I}_{1,2|34}$]{%
    \includegraphics[height=23mm]{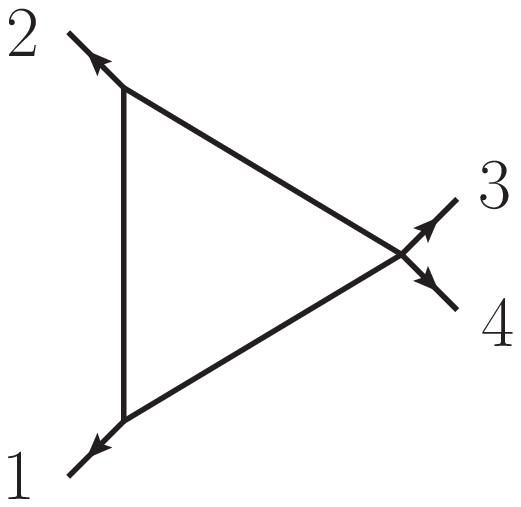}
    \label{fig:2}
  }
  \caption{Integrals contributing to the one-loop  amplitudes.}
  \label{fig:1_2}
\end{figure}

One loop amplitudes may be evaluated though a color-dressed generalization of the 
strategy applicable to all supersymmetric field theories \cite{Britto:2005ha}, by evaluating 
first the coefficients of box integrals through quadruple cuts, then those of triangle integrals
through triple cuts and last the coefficient of bubble integrals, which should vanish in 
our case up to use of eq.~(\ref{vanishingbeta}). Alternatively, they may be constructed 
only from their color-dressed two-particle cuts.
As mentioned in (\ref{separate}), we will  separate the $\NeqFour$ part of the amplitude as
\be
{\cal A}_4^{\beta,(1)}={\cal A}_4^{\NeqFour,(1)}+{\cal A}_4^{\text{extra},(1)} \ ,
\ee
where ${\cal A}_4^{\NeqFour,(1)}$ is well-known \cite{Green:1982sw}:
\be
{\cal A}_4^{\NeqFour,(1)} &=& 
 -\gym^2\,s_{12}s_{23}{\cal A}_4^{\NeqFour,(0)}\Big[N\Big(
\mathcal{I}_{1,2,3,4} (\Tr_{1234}+\Tr_{4321})
+\mathcal{I}_{1,3,4,2}(\Tr_{1342}+\Tr_{2431})
\cr
&&\qquad\qquad\qquad\qquad\qquad
+\mathcal{I}_{1,4,2,3}(\Tr_{1423}+\Tr_{3241})\Big)
\nonumber\\[2pt]
&&\qquad
+2\,\left(\mathcal{I}_{1,2,3,4}+\mathcal{I}_{1,4,2,3}+\mathcal{I}_{1,3,4,2}\right)\left(\Tr_{12}\Tr_{34}+\Tr_{13}\Tr_{24}+\Tr_{14}\Tr_{23}\right)\Big]
\ee
with ${\cal I}_{1,2,3,4}$ the scalar box integral shown in fig.~\ref{fig:1}
\be
{\cal I}_{1,2,3,4}=-i\int \frac{d^dl}{(2\pi)^d}\frac{1}{l^2(l-k_1)^2(l-k_1-k_2)^2(l+k_4)^2} \ .
\ee 
The terms included in ${\cal A}_4^{\text{extra},(1)}$ have a more complicated structure which we now discuss.

\subsection{
$\beta$-dependent one-loop corrections to four-point amplitudes in trace basis
\label{explicit_1loop}}

As explained in \S~\ref{susy_Ward}, amplitudes may be organized following the number 
of external lines in the vector multiplet (\ref{fieldconfig}). 
Moreover, due to minimal supersymmetry, color-stripped amplitudes
related by non-cyclic permutations of external legs are independent. For each 
configuration of 
external ${\cal N}=1$ multiplets we will choose one representative and list here the 
corresponding complete one-loop amplitudes; each color-stripped sub-amplitude
will be expressed as a sum of box and triangle integrals with the appropriate 
momentum-dependent coefficients. 

\

\noindent{$\bullet$ Four-gluon amplitude:} As follows from the discussion in previous sections,
this amplitude is not modified in the presence of the deformation:
\begin{equation}
\mathcal{A}_4^{\text{extra},(1)}(1g^{+},2g^{-},3g^{+},4g^{-})=0 \ .
\end{equation}
Moreover, R-charge conservation requires that the four-point three-gluon amplitude 
vanishes identically.

\

\noindent
{$\bullet$ Two-gluon two-scalar amplitude:} As discussed in \S~\ref{allorders}, 
the first deformation-dependent correction to this amplitude has a double-trace structure and
involves a nontrivial $U(1)$-charge flow between the fields in the two traces. 
At higher orders in the $1/N$ expansion other trace structures are deformed as well.
Since such terms do not have a tree-level counterpart, the expected structure of IR 
divergences~\cite{Akhoury:1978vq}-\cite{Aybat:2006mz} requires that they be finite. 
This is indeed the case; the deformation-dependent terms are  
\begin{eqnarray}
\label{Aggss}
&&\mathcal{A}^{\text{extra},(1)}_4(1g^+,2g^-,3\phi^{14},4\phi^{23})=
 h(\gym, N, q)^{2}(q-q^{-1})^{2}\;\frac{\langle 23\rangle^2}{\langle 13\rangle^2}\;\B(1234)\\
&&\qquad
\times\left[\left(1-\frac{2}{N^{2}}\right)(\Tr_{13}\Tr_{24}+\Tr_{14}\Tr_{23})
   -\frac{1}{N}(\Tr_{1324}+\Tr_{1432})-\frac{2}{N^{2}}\Tr_{12}\Tr_{34}\right] \ .
\nonumber
\end{eqnarray}
Here the overall factor $h(\gym, N, q)^2$ is given by the solution to the 
equation (\ref{vanishingbeta}). The terms of higher order in $1/N$ originate
from the tree-level double-trace amplitudes. 
The momentum-dependent function $\B(1234)$ is a particular combination 
of box and triangle integrals (see fig.~\ref{fig:1_2} for notation)
\begin{equation}
\B(1234)= 
s_{12}s_{14}\mathcal{I}_{1,2,3,4}
-s_{12}(\mathcal{I}_{12|3,4}+\mathcal{I}_{34|1,2})
-s_{14}(\mathcal{I}_{14|2,3}+\mathcal{I}_{23|1,4}) \ .
\label{6dbox}
\end{equation}
This expression may be written as a box integral with a Gram determinant numerator; 
one may also recognize it as the  six-dimensional scalar box integral. 
Evaluating $B(1234)$ (either using the known expressions for the contributing 
integrals or by recognizing it as a six-dimensional integral  \cite{Bern:1993kr}) leads to 
\be
\B(1234)=-\frac{1}{(4\pi)^2}\left[\left(\ln \frac{s_{12}}{s_{23}}\right)^2+\pi^2\right] \ ,
\ee
which is indeed finite in the IR, as expected.

\

\noindent{$\bullet$ One-gluon one-scalar two-fermion amplitude:} The single-trace terms
in this amplitude acquire, as at tree-level, $\beta$-dependent factors which depend on the 
specific color structure. Moreover, since any pair of fields carries nontrivial R-charge, all 
double-trace terms will receive $\beta$-dependent corrections. Because of this it is convenient
to quote the complete amplitude rather than just ${\cal A}_4^{\text{extra},(1)}$:
\begin{eqnarray}
& &\mathcal{A}_4^{\beta,(1)}(1g^{+},2\phi^{i4},3\psi^{j},4\psi^{k})  
=   \gym h(\gym, N, q)\; s_{12}s_{23} \; 
\mathcal{A}_4^{\beta,(0)}(1g^{+},2\phi^{i4},3f^{j},4f^{k})\Big[
\\
&&\qquad \qquad 
   \times N\left(
            \mathcal{I}_{1,2,3,4}\left[q\,\Tr_{1234}+q^{-1}\,\Tr_{1432}\right]
        +  \mathcal{I}_{1,2,4,3}\left[q^{-1}\,\Tr_{1243}+q\,\Tr_{1342}\right] \right.\cr
&&\left.\qquad \qquad \qquad \qquad \qquad \qquad \qquad \qquad \qquad  \qquad \qquad        
        + \mathcal{I}_{1,3,2,4}\left[q^{-1}\,\Tr_{1324}+q\,\Tr_{1423}\right]\right)\cr
 &  & \qquad \qquad\quad
 + (q+q^{-1})\left(\mathcal{I}_{1,2,3,4}+\mathcal{I}_{1,2,4,3}
 +\mathcal{I}_{1,3,2,4}\right)\left[\Tr_{14}\Tr_{23}+\Tr_{13}\Tr_{24}+\Tr_{12}\Tr_{34}\right]
 \Big] \ .
 \nonumber
\end{eqnarray}
The double-trace $1/\epsilon^2$ poles vanish identically, as in the 
underformed theory, as required by the absence of a tree-level double-trace amplitude 
with only one field in the vector multiplet~(\ref{treegsff}). The remaining double-trace 
$1/\epsilon$ terms introduce $\beta$-dependence in the soft anomalous dimension matrix.
\

There are two independent four-point 
amplitudes with four fields in the chiral multiplet: those in which 
all fields belong to one multiplet and its conjugate 
and amplitudes in which the fields  belong to two different 
chiral multiplets. We will choose
$\mathcal{A}(1\phi^{23},2\phi^{14},3\phi^{23},4\phi^{14})$
and
$\mathcal{A}(1\phi^{23},2\phi^{14},3\phi^{13},4\phi^{24})$
as representatives of these two cases, respectively.

\noindent{$\bullet$ like-charge four-scalar amplitude:} At ${\cal O}(1/N)$ these amplitudes 
are rather constrained; for the amplitude $\mathcal{A}(1\phi^{23},2\phi^{14},3\phi^{23},
4\phi^{14})$ the discussion in \S~\ref{allorders} allows a nonzero value 
only for the $\Tr_{13}\Tr_{24}$ structure. At higher orders in $1/N$ however all trace 
structures  may be corrected. We find that
\begin{eqnarray}
\label{samescalars}
\mathcal{A}^{\text{extra},(1)}(1234)&=& \gym^2
\Big[\;N\Big(-1+\h^{4}+\frac{4\h^{4}}{N^{2}}(q-q^{-1})^{2}\Big)\,[\Tr_{1234}+\Tr_{1432}]\\
&&\qquad \;\,\,
-\frac{1}{N}\h^{4}(q-q^{-1})^{2}(q^{2}+q^{-2})\,[\Tr_{1324}+\Tr_{1423}+\Tr_{1243}+\Tr_{1342}]
\nonumber\\
&&\quad
+\Big(-2+2\h^{4}+\frac{1}{N^2}\h^{4}(q-q^{-1})^{4}\Big)\,
[\Tr_{14}\Tr_{23}+\Tr_{12}\Tr_{34}]\nonumber\\
&&\qquad\qquad \;
+\big(-2+\h^{4}(q^{4}+q^{-4})\big)\;\Tr_{13}\Tr_{24}\Big] \; \B(1234) \ ,
\nonumber
\end{eqnarray}
where $\h=\gym^{-1}h(\gym, N, q)$ and $\B(1234)$ is defined in (\ref{6dbox}). 
Using (\ref{vanishingbeta}) it is easy to see 
that the first and third line are ${\cal O}(1/N)$ and ${\cal O}(1/N^2)$ respectively, 
as required by the discussion in \S~\ref{allorders}. Since these terms are proportional to the
combination of integrals in eq.~(\ref{6dbox}) these corrections are IR-finite, as required by the 
analysis in \S~\ref{examples}.

\smallskip

\noindent{$\bullet$ different-charge four-scalar amplitude:} While only two two-trace structures
are corrected at ${\cal O}(1/N)$, all trace structures are corrected at higher orders. For the field
configuration $\mathcal{A}(1\phi^{23},2\phi^{14},3\phi^{13},4\phi^{24})$, the various 
color-ordered terms are as follows:
\begin{eqnarray}
 {A}_{4;2}^{\text{extra},(1)}(14;23)&=& \gym^2\Big(-1+\frac{1}{2}\h^{4}(q^{4}+q^{-4})+\frac{1}{2N^{2}}
\h^{4}(q-q^{-1})^{4}\Big)\;\B(1234)\cr
&&
+\gym^2\big(-2+\h^{2}(q^{2}+q^{-2})\big)(s_{12}s_{13}\mathcal{I}_{1,2,4,3}-s_{14}s_{13}\mathcal{I}_{1,3,2,4})
\\[2pt]
 {A}_{4;2}^{\text{extra},(1)}(13;24)&=& \gym^2(-1+\h^{4})\B(1234)
 \cr
 &+&\gym^2\big(-2+\h^{2}(q^{2}+q^{-2})\big)\mathcal{I}_{1,2,4,3}+2\gym^2(1-\h^{2})\mathcal{I}_{1,3,2,4}\\[2pt]
 {A}_{4;2}^{\text{extra},(1)}(12;34)&=& \gym^2\Big(-1+\h^{4}+\frac{1}{2N^{2}}\h^{4}(q-q^{-1})^{4}\Big)\;\B(1234)\cr
&&\quad
+2\gym^2(-1+\h^{2})(s_{12}s_{13}\mathcal{I}_{1,2,4,3}-s_{14}s_{13}\mathcal{I}_{1,3,2,4})\\
 {A}_{4;1}^{\text{extra},(1)}(1234)&=& {A}_{4;1}^{\text{extra},(1)}(1432)
 = \frac{\gym^2}{2}N\Big(-1+\h^{4}
-\frac{1}{N^{2}}\h^{4}(q-q^{-1})^{4}\Big)\; \B(1234)\\
 {A}_{4;1}^{\text{extra},(1)}(1243)&=& {A}_{4;1}^{\text{extra},(1)}(1342)
 \cr
& =& \frac{\gym^2}{N}\h^{4}(q-q^{-1})^{2}\;\B(1234)
+ \frac{\gym^2}{N}\h^{2}(q-q^{-1})^{2}s_{13}s_{14}\mathcal{I}_{1,3,2,4}\\
 {A}_{4;1}^{\text{extra},(1)}(1324)&=& {A}_{4;1}^{\text{extra},(1)}(1423)^{*}=
- \frac{\gym^2}{N}\h^{4}(q-q^{-1})^{2}q^{2}\;\B(1234)
\\
&&\!\!\!\!\!\!\!\!\!\!\!\!\!\!\!\!\!\!\!\!\!\!\!\!\!\!\!\!\!\!\!\!\!\!\!\!\!\!\!\!\!
- \frac{\gym^2}{N}\h^{2}(q-q^{-1})^{2}s_{12}s_{13}\mathcal{I}_{1,2,4,3}
+ \frac{\gym^2}{N}\Big(N^{2}(1-\h^{2}q^{2})+\h^{2}(q-q^{-1})^{2}\Big)s_{13}s_{14}\mathcal{I}_{1,3,2,4}
\nonumber
\label{55}
\end{eqnarray}
where, as before, $\h=\gym^{-1}h(\gym, N, q)$.
The ${\cal O}(N)$ term in eq.~(\ref{55}) combines with the undeformed amplitude 
to produce the $\beta$-dependent phase factor appropriate for a single-trace planar 
amplitude. The charge assignment forbids similar phase factors for the other 
single-trace  amplitudes.
All triangle integrals may be combined in the six-dimensional box integral $\B(1234)$.
It is not difficult to see that a leading IR divergence exists, as expected, only for
trace structures that have a tree-level counterpart, cf. \S~\ref{examples}.

\subsection{One-loop amplitudes in the modified structure constant basis}

The results described above may also be organized in terms of the modified 
structure constants (\ref{fbeta}). The resulting expressions may also be obtained 
directly in this basis by using tree-level amplitudes dressed with the structure 
constants $f_\beta$ in unitarity cuts. Similarly to the undeformed case,  some of 
the complexity of the expressions in \S~\ref{explicit_1loop} is absorbed in the 
contraction of structure constants.

An inspection of the unitarity cuts and of the tree-level amplitudes in this 
representation reveals that five combinations of modified structure constants 
can appear:
\be
\label{kappa1}
\kappa_\beta(1234;q)&=&{h}(\gym,N,q)f^{a_1 bc}f^{a_2 cd}f^{a_3 de}f_{\beta}^{a_4 ea}\\
\kappa_0(1234;q)&=&{h}(\gym,N,q)^2f^{a_1 bc}f^{a_2 cd}f_\beta^{a_3 de}f_{-\beta}^{a_4 ea}\\
\kappa_1(1234;q)&=&{h}(\gym,N,q)^4f_\beta^{a_1 bc}f_{-\beta}^{a_2 cd}f_{-\beta}^{a_3 de}f_{\beta}^{a_4 ea}\\
\kappa_2(1243;q)&=&\kappa_0(1243;q^{-1})=
{h}(\gym,N,q)^2f^{a_1 bc}f^{a_2 cd}f_{-\beta}^{a_4 de}f_{\beta}^{a_3 ea}
\\
\kappa_3(1324;q)&=&{h}(\gym,N,q)^2f^{a_1 bc}f^{a_3 cd}f_\beta^{a_2 de}f_{\beta}^{a_4 ea} ~ ,
\label{kappa5}
\ee
where ${h}(\gym,N,q)$ is the coefficient of the superpotential which renders the theory UV-finite, given 
by the solution to eq.~(\ref{vanishingbeta}) at one- and two-loop order. 
Other products of structure constants not involving the deformation parameter also appear.
Below we will focus on the $\beta$-dependent terms which will depend on differences 
between various $\kappa$ factors for generic $q$ and for $q=1$.
While in general the translation of products of structure constants to the trace basis does not 
involve terms proportional to $N^{-2}$, such terms nevertheless appear due to the structure of 
${h}(\gym,N,q)$.
For example, one may check that 
\be
&&\kappa_0(1234;q)={h}(\gym,N,q)^2\Big[\;2\Tr_{12}\Tr_{34}
+N(\Tr_{1234}+\Tr_{1432})\\
&+&(q+q^{-1})(\Tr_{13}\Tr_{24}+\Tr_{14}\Tr_{23})+\frac{1}{N}(q-q^{-1})^2(\Tr_{1234}-\Tr_{1324}-\Tr_{1423}+\Tr_{1432}) \Big]\ ,
\nonumber
\ee
implying that
\be
\mathcal{A}^{\text{extra},(1)}_4(1g^+,2g^-,3\phi^{23},4\phi^{14})=
 \left(\kappa_0(1234;q)-\kappa_0(1234;1)\right)\B(1234) \ .
\label{ggss}
\ee
The $q$ dependence of ${h}(\gym,N,q)$ participates nontrivially in this relation.

One may similarly show that the other one-loop amplitudes evaluated in the previous section
can be written as
\be
\mathcal{A}_4^{\beta,(1)}(1g^{+},2\phi^{i4},3\psi^{j},4\psi^{k})  &=&  \gym
s_{12}s_{23}A^{(0)}_4(1234)\Big(\kappa_\beta(1234;q)\, {\cal I}^4_{1,2,3,4}\cr
&&+\kappa_\beta(1423;q)\, {\cal I}^4_{1,4,2,3}
+\kappa_\beta(1342;q)\, {\cal I}^4_{1,3,4,2}\Big) \ ,
\label{gsff}
\\[2pt]
\mathcal{A}^{\text{extra},(1)}(1\phi^{23},2\phi^{14},3\phi^{23},4\phi^{14})
&=&\frac{1}{2}\left(\kappa_0(1234;q)-\kappa_0(1234;1)\right)\B(1234) \ ,
\label{ssss}
\\[2pt]
\mathcal{A}^{\text{extra},(1)}(1\phi^{23},2\phi^{14},3\phi^{13},4\phi^{24})&=&
 \gym^{-2}\left(\kappa_1(1234;q)-\kappa_1(1234;1)\right)\,\B(1234)\cr
&+& \left(\kappa_2(1243;q)-\kappa_2(1243;1)\right)\,s_{12}s_{24}\, {\cal I}^{4}_{1,2,4,3}\cr
&+& \left(\kappa_3(1324;q)-\kappa_3(1324;1)\right)\,s_{13}s_{23}\, {\cal I}^{4}_{1,3,2,4} \ .
\ee

The $q$ factors dressing each color structure depend on the charges of virtual particles;
it turns out that, in the amplitudes in eqs.~(\ref{ggss}) and (\ref{ssss}), the internal charge  
configuration is symmetric. One may use the identity
\be
\kappa_0(1234;q)-\kappa_0(1234;1)\mapsto \frac{1}{2}
\left(\kappa_0(1234;q)+\kappa_0(1234;q^{-1})-2\kappa_0(1234;1)\right)
\ee
to make this symmetry manifest. Similar identities do not seem to exist for the other combinations of 
structure constants, which is consistent with the internal charge configurations for the different-charge 
scalar amplitudes not being symmetric.

With the one-loop amplitudes written in this color basis it is not difficult to search for the color/kinematic 
duality. The amplitude in eq.~(\ref{gsff}), whose color factors obey the Jacobi identity, may be easily 
seen to indeed exhibit it: similarly to the undeformed theory, Jacobi transformations map the three 
terms into each other. 
Other amplitudes however, such as  (\ref{ggss}), are not invariant under Jacobi transformations. One may 
trace this both to the absence of Jacobi identities for color factors as well as to the presence of both box and 
triangle integrals with the same color factor.

\subsection{The symmetries of the one-loop four-point amplitudes}

Similarly to $\NeqFour$ one-loop amplitudes, the one-loop amplitudes of the $\beta$-deformed
theory have certain symmetries, in some cases only after certain state-dependent rational functions
of spinor products are extracted. Some of these symmetries are dictated by the configuration
of external legs. For example, one may expect that 
${\cal A}^{\beta}_4(1\phi^{jk},2\phi^{i4},3\phi^{jk},4\phi^{i4})$ with $i\ne j\ne k = 1,2,3$
({\it i.e.} an amplitude with all external states carrying the same type of $U(1)$ charge) 
is invariant under conjugation together with a certain relabeling of momenta (the color factors are not relabeled): 
\be
\label{Csym}
\C: ~\langle\rangle\leftrightarrow []\quad\{1,2,3,4\}\mapsto \{2,1,4,3\}  
\qquad
\text{or}
\qquad
\C': ~\langle\rangle\leftrightarrow []\quad\{1,2,3,4\}\mapsto \{4,3,2,1\}  \ .
\ee
It is not difficult to check that this is indeed a symmetry of eq.~(\ref{samescalars}). 
This amplitude has further symmetries stemming from its external states being identical 
in pairs; one may check that, indeed, eq.~(\ref{samescalars}) is invariant 
under the relabeling
\be
1\leftrightarrow 3 
\qquad
\text{or}
\qquad
2\leftrightarrow 4 \ . 
\ee
For some of the single-trace components, these symmetries may be also identified 
as a consequence of the inversion symmetry of scattering amplitudes: 
$A(1,\dots,n)=(-1)^nA(n,\dots,1)$. For the double-trace terms these transformations
are symmetries of the color structures, which suggests that they should also be
symmetries of the kinematic part.

Similarly, the two-gluon two-scalar amplitude $\mathcal{A}(1g^{+},2g^{-},3\phi^{ij},4\phi^{kl})$ 
($i\ne j\ne k \ne l$)
has two independent symmetries. One of them, present also in the tree-level amplitude, 
is the $\C$ transformation in eq.~(\ref{Csym}).
The second one is a symmetry of the amplitude only after the factor 
$\langle 23\rangle^2/\langle 13\rangle^2$ is stripped off; by inspecting the remainder  
it is not difficult to see that it is invariant under  
\be
\label{UUpsym}
\U: \{1,2,3,4\}\mapsto\{4,3,2,1\} 
\qquad\qquad
\U': \{1,2,3,4\}\mapsto\{3,4,1,2\} \ .
\ee
It is not difficult to see that $\U'=\U\C$ and $\U=\U'\C$.  
While these are symmetries of amplitudes with identical external states (being in fact a 
simple inversion (U) or a shift transformation by two units (\U')), it is not clear why it 
should survive in the presence of the deformation, when the external states belong to 
different multiplets. This feature suggests that certain four-point amplitudes
exhibit more supersymmetry than the complete theory\footnote{The structure of the 
specific amplitude discussed here is as if all fields may belong to an ${\cal N}=2$ vector 
multiplet.}.
Technically, this symmetry arises because for chargeless and like-charge external states 
the internal states are symmetric under the exchange of internal states with charges 
different from that of the external one.
As we will see, the leading two-loop double-trace terms exhibit this symmetry as well. An 
analysis of cuts of higher loop amplitudes shows that this symmetry may survive at 
two loop level and beyond.

\section{The leading double-trace correction to a two-loop amplitude  \label{twoloop}  }

The UV properties of the one-loop amplitudes constructed in the previous section 
are in line with the expectations based on the finiteness of the theory and the 
consequences of $\NeqOne$ supersymmetry. Quite generally however, one-loop amplitudes 
do not always accurately reflect the properties of higher-loop amplitudes. For example, the 
one-loop amplitudes of the $\NeqFour$ sYM theory do not follow the same pattern as higher-loop 
amplitudes and diverge in $D_c=8$ dimensions and not in $D_c=10$ as suggested by the 
usual expression of the critical dimension \cite{Howe:2002ui, Bern:1998ug}
\be
D_c^{\NeqFour}=4+\frac{6}{L} \ .
\ee
While one-loop amplitudes follow the pattern of critical dimensions suggested by $\NeqOne$ 
supersymmetry, it is nevertheless interesting to explore whether some unexpected 
features emerge in the $\beta$-deformed theory as well. As we will see, this is indeed the case.

\subsection{Generalities}

We will consider here in detail the four-point amplitudes with 
two fields in the vector multiplet and focus on the leading terms arising due to the presence 
of the deformation -- the double-trace terms in this amplitude. We will organized them as
\be
\label{2loopamp}
\mathcal{A}^{\beta,(2)}_4(1g^{+},2g^{-},3\phi^{i4},4\phi^{jk})&=&
\mathcal{A}^{\NeqFour,(2)}_4(1234)+\mathcal{A}^{\text{extra},(2)}_4(1234)
\\
\mathcal{A}^{\text{extra},(2)}_4(1234)&=&
(q^{2}+q^{-2}-2)\frac{\langle23\rangle^{2}}{\langle13\rangle^{2}}
\mathcal{M}^{\text{extra},(2)}_4(1234) \ ,
\ee
where $i\ne j \ne k = 1,2,3$. For the field configuration discussed here, the term 
proportional to $\Tr_{12}\Tr_{34}$ is absent as there is no $U(1)$-charge flow 
between fields in the two trace factors.  As discussed in \S~\ref{allorders}, 
the corresponding correction vanishes identically. Thus, the first $\beta$-dependent 
correction to the scalar factor has the form: 
\be
\mathcal{M}^{\text{extra},(2)}_4(1234)=
{M}^{\text{extra},(2)}_{13;24}\Tr_{13}\Tr_{24}
+
{M}^{\text{extra},(2)}_{14;23}\Tr_{14}\Tr_{23} \ .
\ee
An expression in terms of products of modified structure constants (\ref{fbeta}) 
should also exist once higher order terms in the $1/N$ expansion are included.

Similarly to the $\NeqFour$ amplitudes, each trace structure may be expressed as a sum
of parent integrals ({\it i.e.} in this case the planar and non-planar massless double-box 
integrals) each of them having a certain kinematic numerator factor:
\begin{equation}
M^{\text{extra},(2)}_{*;*}=\sum_{i}c_{i}I_{i} \ .
\end{equation}
The numerator factors $c_i$ contain inverse propagators; they cancel some of the 
propagators appearing in the parent graphs, thus leading to contact terms. As discussed 
at length in \cite{Bern:2008pv, Bern:2010tq}, there is no unique assignment 
of contact terms to parent graphs; for this reason we will write out explicitly the contact 
terms. 
Even more so than in $\NeqFour$ sYM theory, the proliferation of contact terms implies 
that there are many superficially different presentations of amplitudes in theories with 
reduced supersymmetry. Symmetries provide
an efficient way of organizing the result. The two-loop amplitude is expected to be invariant
under the $\C$ transformation in eq.~(\ref{Csym});
in fact, the spinor ratio factor and  the scalar function ${\cal M}$ are separately invariant 
under this symmetry. 

It is possible to see that the transformations (\ref{UUpsym}) are symmetries of the 
generalized cuts of the two-loop contribution to the scalar function ${\cal M}$. 
Moreover, the cuts of ${M}^{\text{extra},(2)}_{14;23}$ can be saturated by 
cuts of integrals whose topologies are manifestly invariant under $\U$ but 
not invariant under $\U'$
while
the cuts of ${M}^{\text{extra},(2)}_{13;24}$ can be saturated by 
cuts of integrals whose topologies are manifestly invariant under $\U'$ but 
not invariant under $\U$.
\footnote{Of course, the non-manifest symmetries map various integrals into each 
other.}  
It is therefore natural to enforce a different manifest symmetry on the kinematic 
numerators in the two components of the scalar function.

\subsection{${M}^{\text{extra},(2)}_{14;23}$ }

\begin{figure}
  \centering
  \subfigure[$I_1$]{%
    \includegraphics[height=20mm]{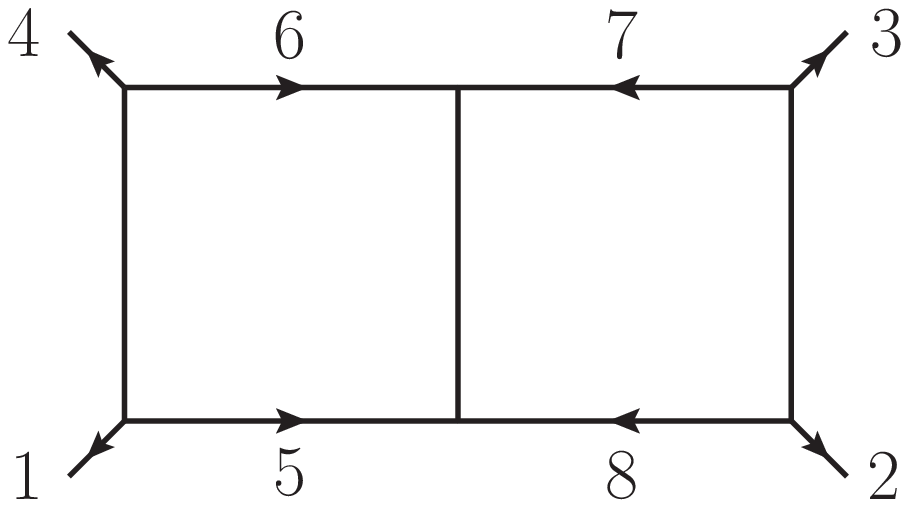}
    \label{fig:4}
  }
  \subfigure[$I_2$]{%
    \includegraphics[height=20mm]{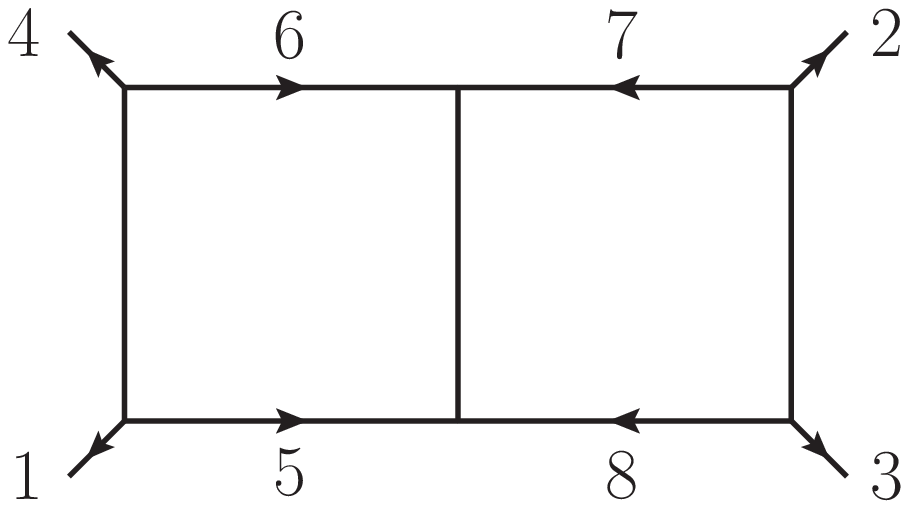}
    \label{fig:5}
  }
  \subfigure[$I_3$]{%
    \includegraphics[height=17.75mm]{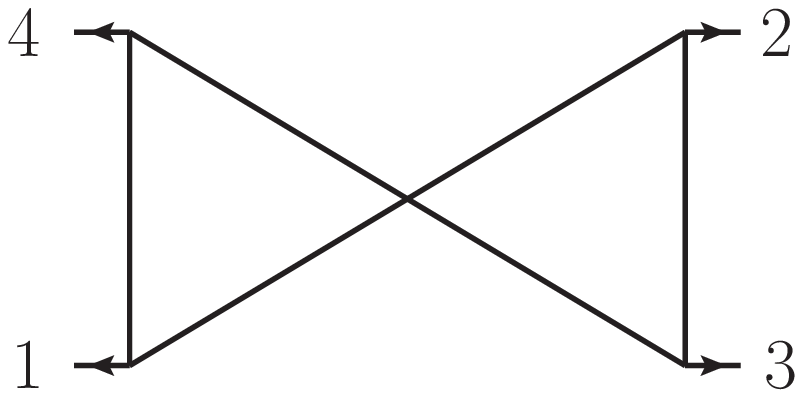}
    \label{fig:6}
  }
  \subfigure[$J_1$]{%
    \includegraphics[height=20mm]{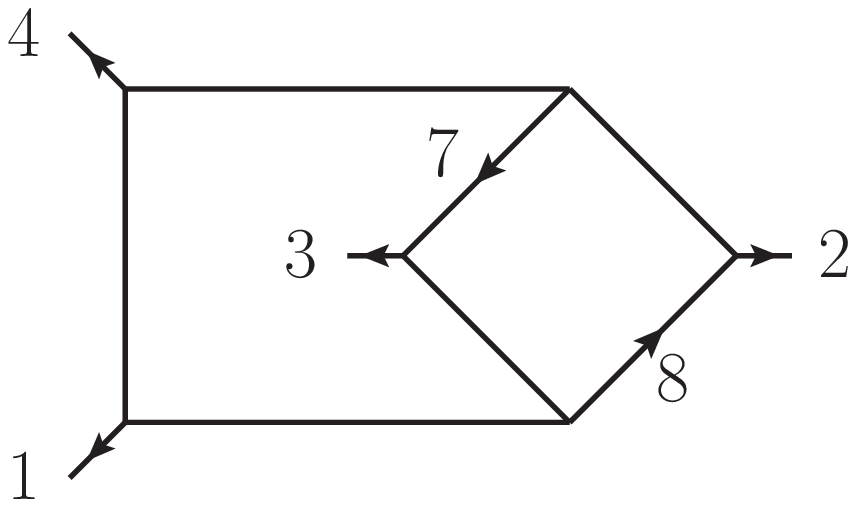}
    \label{fig:7}
  }
  \subfigure[$J_2$]{%
    \includegraphics[height=17.5mm]{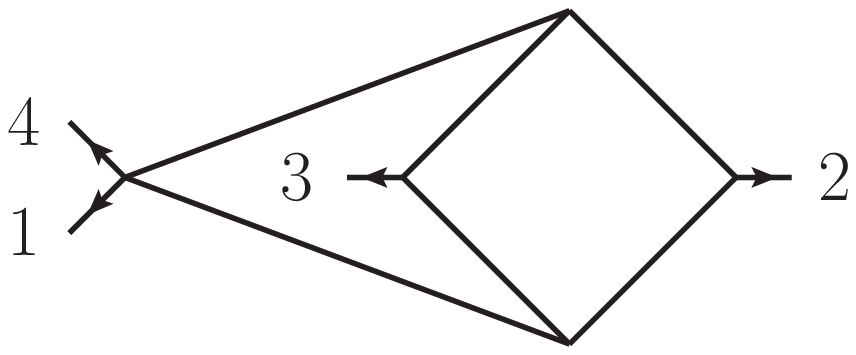}
    \label{fig:8}
  }
  \subfigure[$J_3$]{%
    \includegraphics[height=20mm]{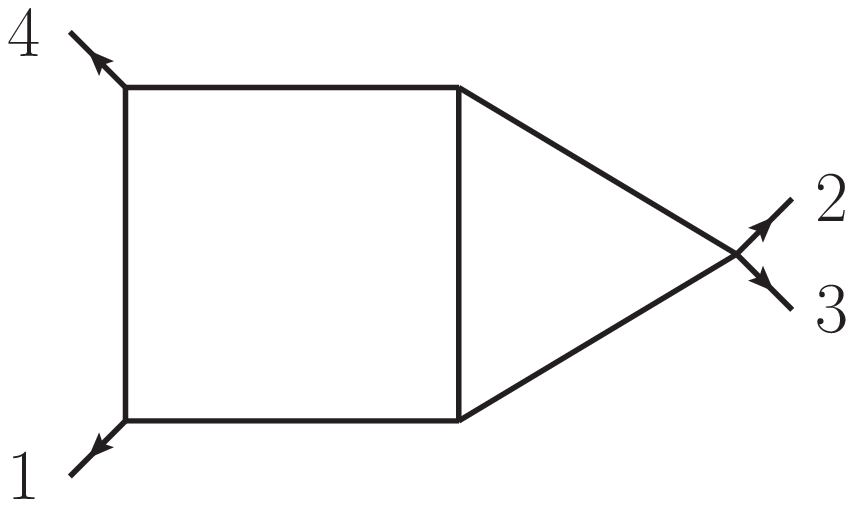}
    \label{fig:9}
  }
  \subfigure[$J_4$]{%
    \includegraphics[height=20mm]{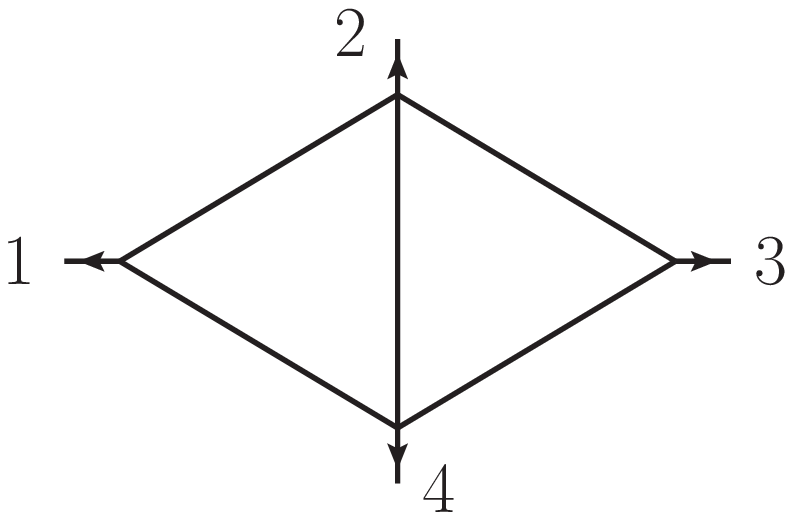}
    \label{fig:10}
  }
  \subfigure[$J_5$]{%
    \includegraphics[height=20mm]{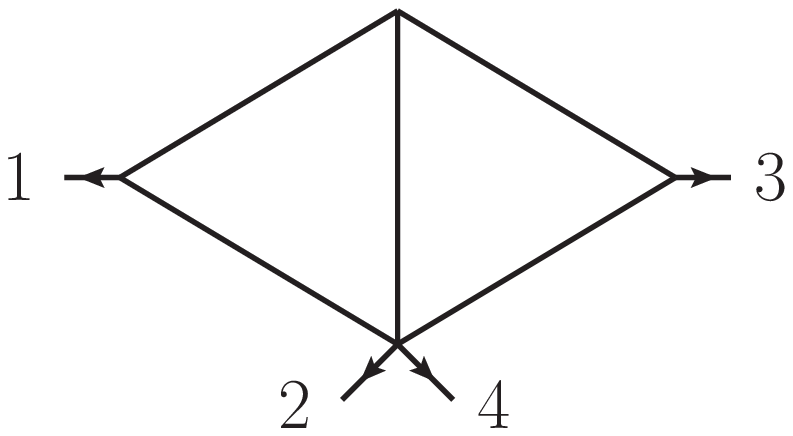}
    \label{fig:11}
  }
  \subfigure[$K_1$]{%
    \includegraphics[height=20mm]{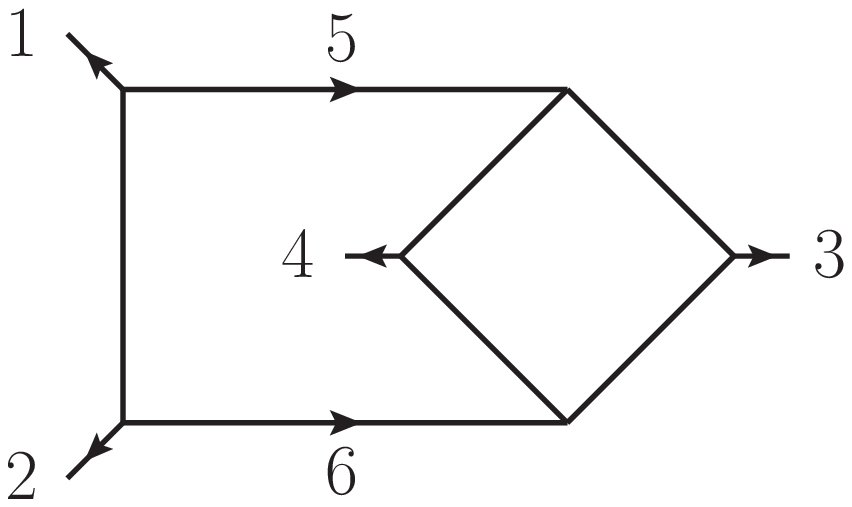}
    \label{fig:12}
  }
  \subfigure[$K_2$]{%
    \includegraphics[height=17.7mm]{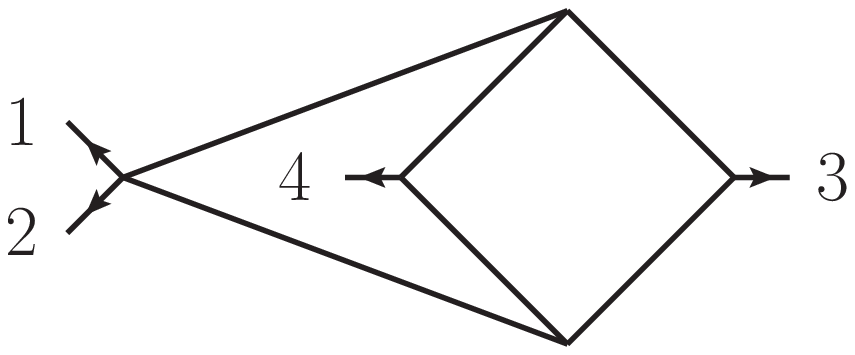}
    \label{fig:13}
  }
  \subfigure[$K_3$]{%
    \includegraphics[height=20mm]{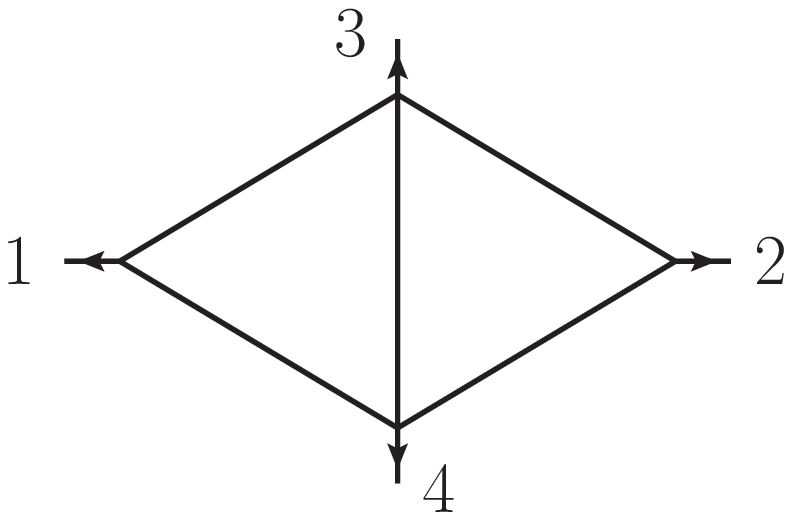}
    \label{fig:14}
  }
  \subfigure[$K_4$]{%
    \includegraphics[height=20mm]{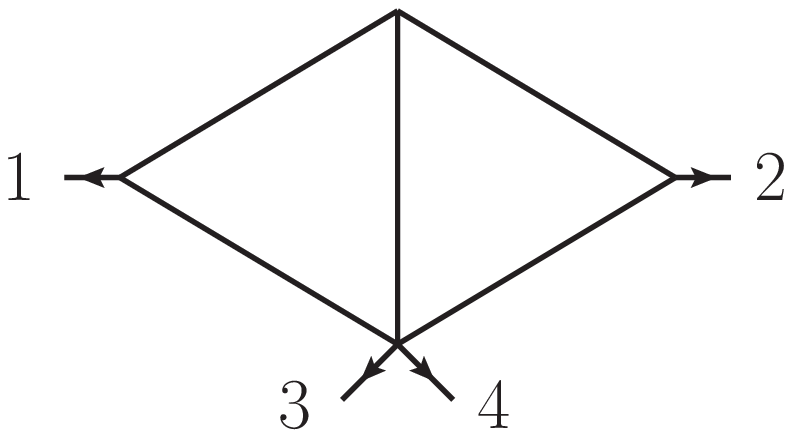}
    \label{fig:15}
  }
  \subfigure[$L_1$]{%
    \includegraphics[height=20mm]{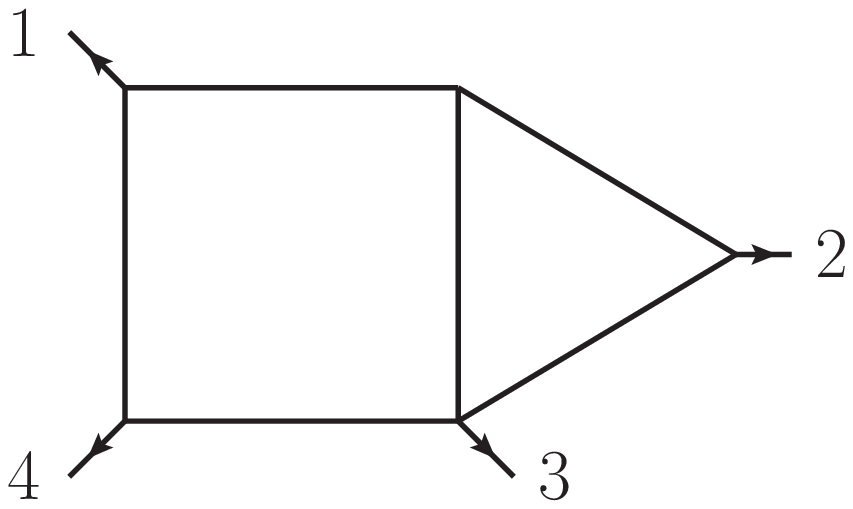}
    \label{fig:16}
  }
  \subfigure[$L_2$]{%
    \includegraphics[height=20mm]{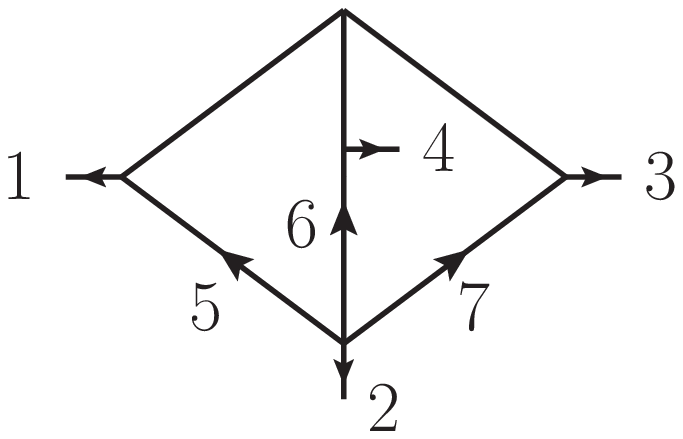}
    \label{fig:17}
  }
  \caption{Integrals contributing to the trace structure $\Tr_{14}\Tr_{23}$ in the 
  two-gluon two-scalar amplitude. Their coefficients are listed in eqs.(\ref{al14_23})-(\ref{de14_23}).}
  \label{fig:4_17}
\end{figure}

The integral topologies entering the $\beta$-dependent contribution to the 
color-stripped amplitude corresponding to the trace structure $\Tr_{14}
\Tr_{23}$ are shown in figure~\ref{fig:4_17}.  
Only the first three are invariant under 
the complete symmetry (\ref{Csym}), (\ref{UUpsym}); some of the integrals for which 
$\C$ is not a symmetry are invariant under either $\U$ or $\U'$ transformations and vice versa. 
We will choose to organize the result in a form that is manifestly invariant under $\C$ 
and $\U$: 
\begin{equation}
M^{\text{extra},(2)}_{14;23}=\sum_{i}\alpha_{i}I_{i}+(1+\C)\sum_{i}\beta_{i}J_{i}
+(1+\U)\sum_{i}\gamma_{i}K_{i}+(1+\U)(1+\C)\sum_{i}\delta_{i}L_{i} \ ,
\end{equation}
with invariance under $\U'$ following from the relation $\U'=\U\C$. 
Given some collection of integrals, there are other ways to make it invariant under 
$\C$ and $\U$; the expression here is singled out by the explicit calculation.

With the definition $\tau_{a,b}=2k_{a}\cdot k_{b}$ ($k_a$ with $a\ge 5$ are loop momenta), 
the numerator factors of the various integrals 
are:
\be
\alpha_{1}&=&
\tau_{1,4}
\left(\tau_{1,8}^{2}+\tau_{2,5}^{2}+\tau_{4,7}^{2}+\tau_{3,6}^{2}+\tau_{1,2}(\tau_{1,8}+\tau_{2,5}+\tau_{4,7}+\tau_{3,6}-2\tau_{1,4})-\tau_{1,8}\tau_{2,5}-\tau_{4,7}\tau_{3,6}\right)
\cr
\alpha_{2}&=&\tau_{1,4}\left(\tau_{1,8}^{2}+\tau_{2,6}^{2}+\tau_{4,7}^{2}+\tau_{3,5}^{2}+\tau_{1,2}(\tau_{1,8}+\tau_{2,6}+\tau_{4,7}+\tau_{3,5})+\tau_{1,8}\tau_{2,6}+\tau_{4,7}\tau_{3,5}\right)
\cr
\alpha_{3}&=&-4\tau_{1,3}\tau_{1,4}
\label{al14_23}
\\[3pt]
\beta_{1}&=&2\tau_{1,4}\left(\tau_{1,8}^{2}+\tau_{4,7}^{2}+\tau_{1,3}(\tau_{1,8}+\tau_{4,7})
\right)
\cr
\beta_{2}&=&-2\tau_{1,4}^{2}
\cr
\beta_{3}&=&2\tau_{1,4}^{2}
\cr
\beta_{4}&=&2\tau_{1,2}
\cr
\beta_{5}&=&\tau_{1,4}-2\tau_{1,2}
\label{be14_23}
\\[3pt]
\gamma_{1}&=&\tau_{1,2}\left(
\tau_{1,5}^{2}+\tau_{2,6}^{2}+\tau_{1,2}(\tau_{1,5}+\tau_{2,6})\right)
\cr
\gamma_{2}&=&2\tau_{1,2}^{2}
\cr
\gamma_{3}&=&2\tau_{1,2}
\cr
\gamma_{4}&=&\tau_{1,3}-\tau_{1,2}
\label{ga14_23}
\\[3pt]
\delta_{1}&=&\tau_{1,4}^{2}
\cr
\delta_{2}&=&-\tau_{1,4}\tau_{2,5}+\textstyle{\frac{1}{2}}\tau_{12}(3\tau_{1,3}+\tau_{2,5}-2\tau_{3,5}+2\tau_{2,7})
\label{de14_23}
\ee
Integration-by-parts identities may be used to reduce some of the integrals with 
nontrivial numerator factors to simpler ones\footnote{We thank Lance Dixon for discussions 
on this point.} along the lines of \cite{Smirnov:1999wz}. We will however not pursue this here.
It is not difficult to see that, integral by integral, ${M}^{\text{extra},(2)}_{14;23}$ 
is UV finite in four dimensions. All the integrals having at least one four-point vertex 
may be obtained (in several ways) by collapsing suitable internal lines in the graphs 
$I_1, I_2, J_1, K_1$ and their images under the symmetries of the amplitude.

\subsection{${M}^{\text{extra},(2)}_{13;24}$ }

\begin{figure}
  \centering
  \subfigure[$J'_1$]{%
    \includegraphics[height=20mm]{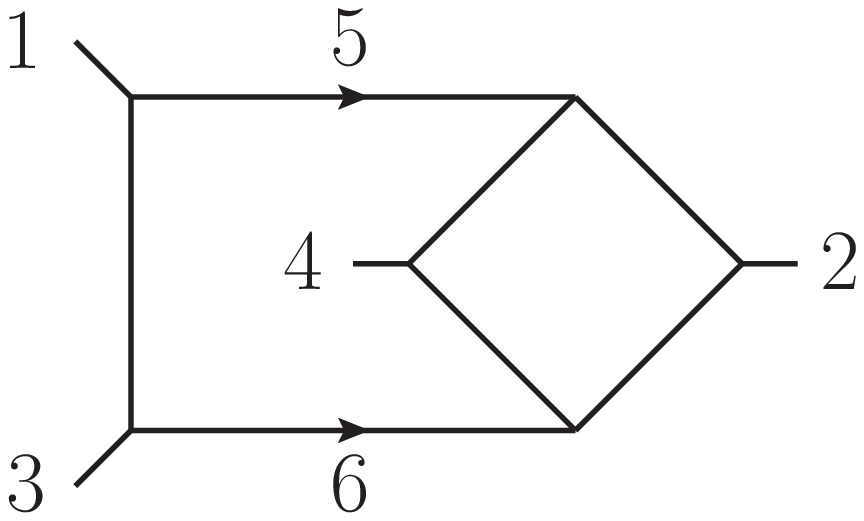}
    \label{fig:18}
  }
  \subfigure[$J'_2$]{%
    \includegraphics[height=20mm]{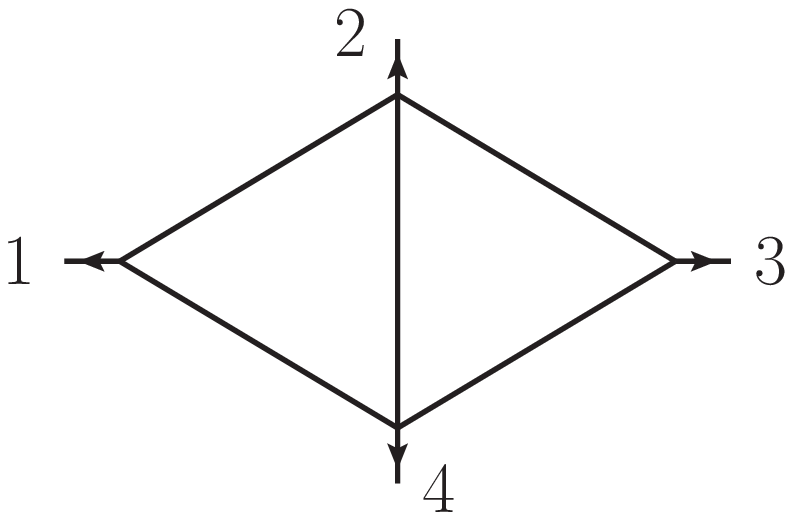}
    \label{fig:19}
  }
  \subfigure[$J'_3$]{%
    \includegraphics[height=20mm]{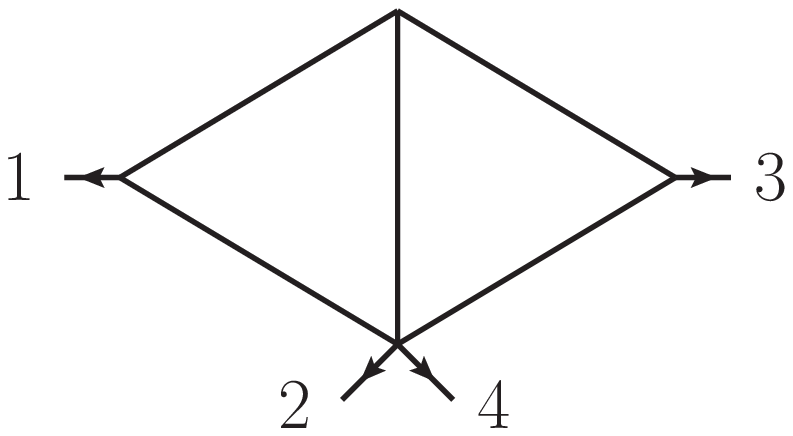}
    \label{fig:26}
  }
  \subfigure[$K'_1$]{%
    \includegraphics[height=20mm]{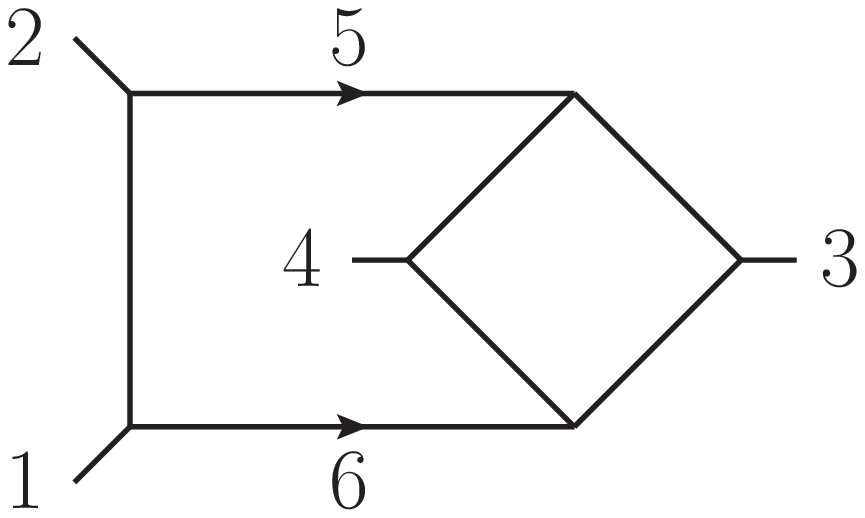}
    \label{fig:21}
  }
  \subfigure[$K'_2$]{%
    \includegraphics[height=17.5mm]{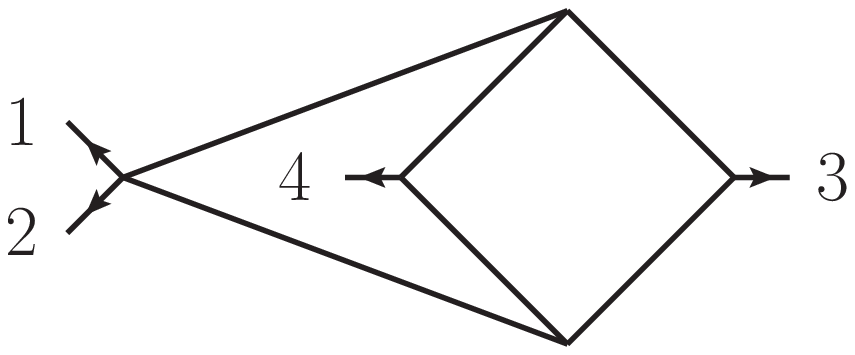}
    \label{fig:22}
  }
  \subfigure[$L'_1$]{%
    \includegraphics[height=20mm]{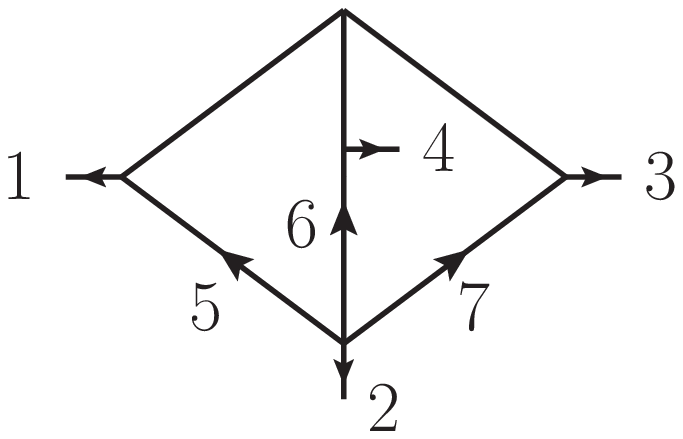}
    \label{fig:24}
  }
  \subfigure[$M'_1$]{%
    \includegraphics[height=20mm]{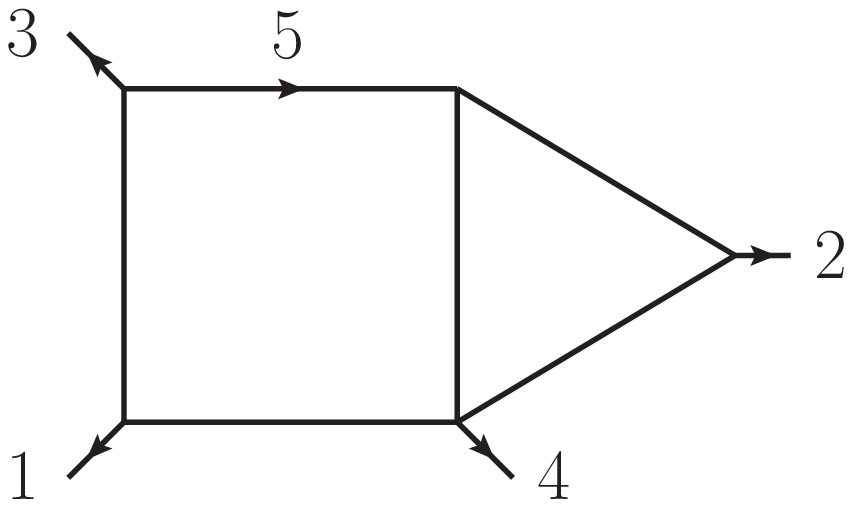}
    \label{fig:25}
  }
  \subfigure[$X'_1$]{%
    \includegraphics[height=20mm]{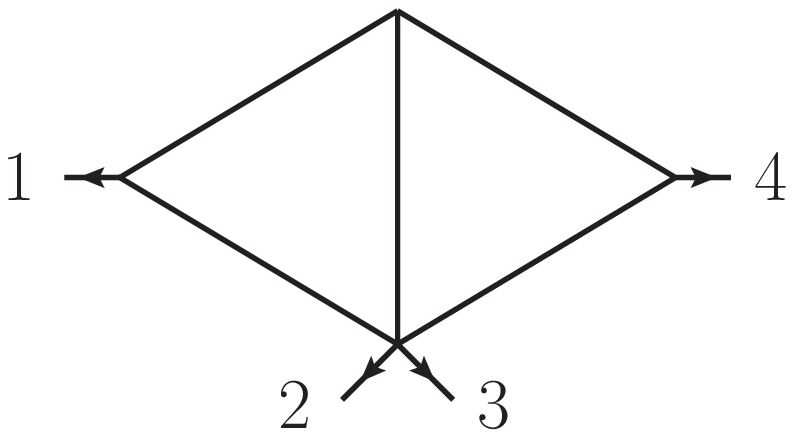}
    \label{fig:20}
  }
  \subfigure[$Y'_1$]{%
    \includegraphics[height=20mm]{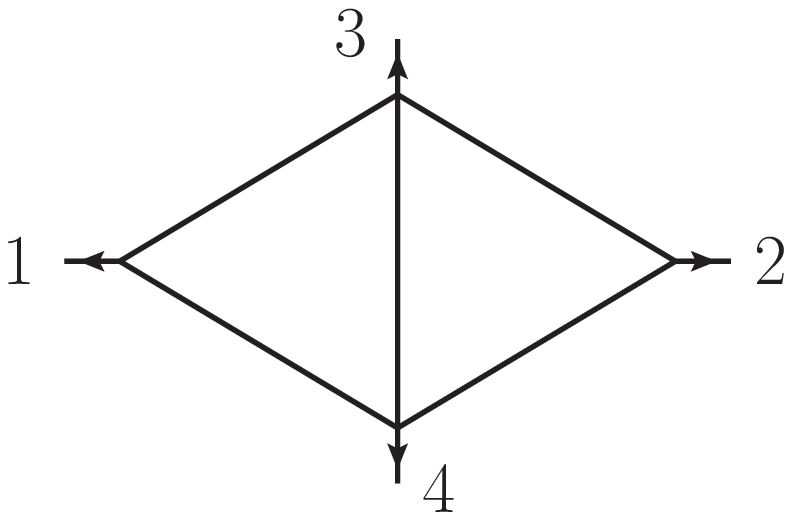}
    \label{fig:23}
  }
  \caption{Integrals contributing to the trace structure $\Tr_{13}\Tr_{24}$ in the 
  two-gluon two-scalar amplitude. their coefficients are listed in eqs.(\ref{bep13})-(\ref{etp1}).}
  \label{fig:18_25}
\end{figure}

The leading deformation-dependent correction to the $\Tr_{13}\Tr_{24}$
trace structure exhibits a larger symmetry than ${M}^{\text{extra},(2)}_{14;23}$; apart from 
invariance under $\C,\,\U$ and $\U'$, it is also invariant under 
\be
{\rm E}:\quad \{1,2,3,4\}\mapsto \{3, 2, 1, 4\} \ .
\ee
Together, these transformations form the symmetry group of 1-loop amplitudes.
The integral topologies entering ${M}^{\text{extra},(2)}_{13;24}$ are shown in 
figure~\ref{fig:18_25}; it turns out that, in this case, it is convenient to choose 
$\C$, $\U'$ and ${\rm E}$ as manifest symmetries:
\begin{eqnarray}
{M}^{\text{extra},(2)}_{13;24}&=&(1+\C)\sum_{i}\beta'_{i}J'_{i}
+(1+\U')\sum_{i}\gamma'_{i}K'_{i}+(1+\U')(1+\C)\sum_{i}\delta'_{i}L'_{i}
\nonumber\\
&+&
(1+\U')(1+\C)(1+\E)\sum_{i}\epsilon'_{i}M'_{i}
+(1+\C)(1+\E)\sum_i\rho'_i X'_i
\nonumber\\
&+&(1+\U')(1+\E)\sum_i\eta'_i Y'_i ~ .
\end{eqnarray}
Similarly to ${M}^{\text{extra},(2)}_{14;23}$, there are other ways to make the 
$\C$, $\U'$ and $\E$ manifest; the expression here follows from the explicit 
calculation; the numerator factor of each integral is given by:
\be
\beta'_{1}&=&
-2\tau_{1,3}\left(\tau_{2,5}\tau_{2,6}+\tau_{4,5}\tau_{4,6}\right)
\cr
\beta'_{2}&=&-2\tau_{1,3}
\cr
\beta'_{3}&=&2\tau_{1,3}
\label{bep13}
\\[3pt]
\gamma'_{1}&=&\tau_{1,2}\left(
\tau_{4,5}^{2}+\tau_{3,6}^{2}+\tau_{1,3}(\tau_{4,5}+\tau_{3,6})\right)
\cr
\gamma'_{2}&=&\tau_{1,2}^{2}
\label{gap13}
\\[3pt]
\delta'_{1}&=&
-\textstyle{\frac{1}{2}}\tau_{1,3}\tau_{2,6}
+
\textstyle{\frac{1}{2}}(\tau_{1,4}-\tau_{1,2})(\tau_{2,5}-\tau_{2,7}+2\tau_{4,5}-2\tau_{4,7})
\label{dep1}
\\[3pt]
\epsilon'_{1}&=&2\tau_{1,3}\tau_{2,5}
\\[3pt]
\rho'_{1}&=&4\tau_{1,3}
\\[3pt]
\eta'_{1}&=&-4\tau_{1,3} ~ .
\label{etp1}
\ee
Similarly to the ${M}^{\text{extra},(2)}_{14;23}$, these corrections are UV finite integral 
by integral in four dimensions. Moreover, as in that case, 
all the integrals having at least one four-point vertex 
may be obtained (in several ways) by collapsing suitable internal lines in the graphs 
$J_1', K_1'$ and $M_1'$ and their images under the symmetries of the amplitude.

\section{The UV 
behavior of one- and two-loop amplitudes}

Simply inspecting the integrals appearing in \S~\ref{oneloop} and \S~\ref{twoloop} it 
is easy to see that they are convergent in four dimensions. 
Unlike $\NeqFour$ sYM, the $\beta$-deformed theory is intrinsically four dimensional
because on the one hand it has minimal supersymmetry and on the other the structure of 
the deformation relies on the existence of three $U(1)$ symmetries.
This makes it difficult to discuss, in parallel with $\NeqFour$ sYM theory, the UV 
convergence properties of the theory in higher  dimensions, since such a theory does 
not exist. 

For fixed dimension $D_0$ the convergence properties of a theory are captured by the degree 
of divergence which counts for each integral the difference between the expected number 
of loop momenta in the numerator and denominator and is a function of the dimensionality 
of space-time $D_0$, number of loops $L$, number of vertices $V$ and propagators $I$.
${\cal N}$-extended supersymmetry improves the naive value of the degree of divergence and, 
for a finite theory, it is expected to lower its value by $2{\cal N}$:
\be
\omega^\text{susy}=\omega^\text{naive}(D_0, L, V, I) - 2{\cal N} \ .
\ee
Even for a theory defined only for some fixed dimension $D_0$, one can still express its UV 
convergence as a formal value of the dimension, $D_0\rightarrow D>D_0$, which sets to zero 
the degree of  divergence. This value is, of course, unphysical since the theory does not exist 
in that dimension. Rather, it corresponds to the naive analytic continuation of the integrals 
composing the amplitudes after all dimension-dependent manipulations have been carried out.
If the theory could be defined in higher dimensions such a continuation does not accurately describe 
the UV properties of the theory because of potential $\mu$-terms ({\it i.e.} terms manifestly proportional 
only to the $(D-4)$ components of loop momenta) which are not evaluated in the original dimension. 
We will formulate in these terms the UV properties of the $\beta$-deformed theory, while keeping in 
mind that the implications of our results are solely for the value of the degree of divergence for $D_0=4$;
we will refer to $D$ as the "formal dimension".


The fact that only scalar box and triangle integrals enter the one-loop amplitude 
implies that the one-loop formal dimension in which the first divergence appears is 
\be
D_c=6 \ .
\ee
This is, in fact, the UV behavior of a generic finite $\NeqOne$ supersymmetric field theory, 
with the slight distinction that the four-point amplitudes do not contain finite combinations
of bubble integrals. 

The situation is more interesting at two-loop level. The five-propagator integrals have 
worse UV behavior; their presence suggests that the analytically-continued 
two-loop amplitude is manifestly finite only for 
\be
D\le 5 \ ;
\ee
this would-be critical dimension is consistent with the behavior of an $\NeqOne$ finite 
theory\footnote{Let us recall that, as mentioned in the introduction, a finite supersymmetric
theory with ${\cal N}$ manifestly-realized supercharges has an expected $L$-loop critical 
formal  dimension $D_c\le 4+2{\cal N}/L$}.

To check whether this UV behavior is improved it is necessary to extract  the leading 
UV-divergent terms in $D=5$. To this end we follow the strategy described in 
\cite{GravityThree, GravityFour, Bern:2010tq}; that is, near the critical dimension 
we notice that the overall UV divergences comes from the integration region in which loop 
momenta are much larger than external momenta, allowing us to capture the leading UV
behavior by expanding in external momenta~\cite{MarcusSagnotti}. 

\begin{figure}
\centering
{\includegraphics[height=28mm]{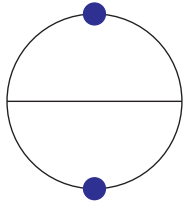}} 
\caption{The unique vacuum integral topology 
capturing the UV properties of the analytically-continued 
$\beta$-dependent double-trace terms in the two-gluon two-scalar amplitude.}
\label{fig:vacuum}
\end{figure}

In the case at hand the relevant integrals to analyze are the 5-propagator ones 
-- $J_4$, $J_5$, $K_3$, $K_4$  for $M_{14;23}^{\beta,(2)}$  and 
    $J'_2$, $J'_3$, $X'_1$, $Y'_1$   for $M_{13;24}^{\beta,(2)}$. Note that there are no 6-
propagator integrals with numerators that are quadratic in loop momenta. 
In the limit in which the external momenta are vanishing all these integrals reduce to a
unique vacuum  topology shown in fig.~\ref{fig:vacuum}. 

Using the symmetries of the numerator factors listed in eqs.~(\ref{al14_23})-(\ref{de14_23}) 
and (\ref{bep13})-(\ref{etp1}) and momentum conservation, the coefficients 
of this vacuum integral in the two scalar functions are as follows:
\be
M_{14;23}^{\beta,(2)}:&&2\beta_4+2\beta_5+2\gamma_3+2\gamma_4 = 0
\\
M_{13;24}^{\beta,(2)}:&&2\beta'_2+2\beta'_3+4\rho'_1+4\eta'_1=0 \ .
\ee
This cancellation implies that the analytic continuation of the leading $\beta$-dependent 
correction to the two-gluon two-scalar amplitude is finite in $D=5$; Lorentz invariance 
implies then that this amplitude first diverges in
\be
D_c = 6 \ .
\ee  
Even though the $\beta$-deformed theory has only ${\cal N}=1$ supersymmetry, as discussed in 
the introduction, this critical formal dimension is characteristic to a finite ${\cal N}=2$ theory. In 
a strict four-dimensional formulation, our result is that the degree of divergence is $\omega=-4$, 
which is smaller by two units than the expected degree of divergence of a finite $\NeqOne$ theory, 
$\omega=-2$.

\section{Summary and further comments \label{conclusion}}
 
In this paper we discussed in detail the scattering amplitudes of the $\beta$-deformed 
$\NeqFour$ sYM theory at the nonplanar level though two loops. The existence of 
double-trace components of tree-level amplitudes makes the scattering matrix of this theory 
unconstrained by the general analysis of \cite{Benincasa:2007xk}; these terms, directly linked 
to the existence of symmetric structure constant couplings as well as to the non-decoupling of 
chiral multiplets charged under the diagonal $U(1)$ subgroup of the gauge group, play 
an important role in the finiteness of the theory. 

We have seen that, apart from planar structures inherited from the $\NeqFour$ theory
due to the structure of the deformation, certain non-planar color structures are also protected 
from being directly affected by the deformation, to all orders in perturbation theory.
These amplitudes may nevertheless depend on the deformation parameter through the
coefficient of the superpotential whose value is fixed in terms of $\beta$, $N$ and the 
gauge coupling by the requirement of conformal invariance.
We have also discussed the loop order at which the known expression for this coefficient 
requires further corrections. We identified a simple color-based argument which recovers 
the form of its one-loop expression and the fact that it also implies that the theory 
is finite at two-loop level. The same argument implies that the next correction 
is required at three-loop level. 
The existence of a superpotential coefficient leading to an all-order finiteness for the
$\beta$-deformed theory is guaranteed by the original argument of 
Leigh and Strassler \cite{Leigh:1995ep}. It should be interesting to find an all-order 
closed-form expression for this coefficient.

By explicitly evaluating four-point loop amplitudes we found that, while at one-loop level
the UV properties of the theory can be explained by its $\NeqOne$ supersymmetry, some 
two-loop amplitudes exhibit an improved behavior. 
It is not clear what is the origin of this improvement. It 
is however interesting to notice its similarity with the improved UV behavior of the 
double-trace terms
in the $\NeqFour$ sYM theory. In that case, a supersymmetry-based explanation of this 
phenomenon was identified in \cite{Berkovits:2009aw}. While we do not understand the 
fundamental origin of our result, an argument  similar to that of \cite{Berkovits:2009aw}
does not appear possible in our case. 
Another explanation of the $\NeqFour$ improved double-trace behavior was suggested 
in \cite{Bern:2010tq}, essentially based on the structure of the amplitude required by
color/kinematics duality. These arguments do not appear to extend in a straightforward way
to the $\beta$-deformed theory due to its different color structure.

We are therefore led to suggest that the improved UV properties of the double-trace 
terms in both the $\beta$-deformed  and in the $\NeqFour$ sYM theory must 
have yet a different explanation, common to both theories, which does not rely
either on supersymmetry or on color/kinematics duality. This also points to further
hidden structure in the non-planar $\NeqFour$ theory.

Naive inspection of the one- and two-loop amplitudes found in \S~\ref{oneloop} and 
\S~\ref{twoloop} does not reveal obvious consequences of the numerator relations 
inherited from the color/kinematics duality of the undeformed theory. 
It is nevertheless possible that loop amplitudes are affected -- and perhaps constrained --
by these tree-level numerator  relations. It would be interesting to identify these 
consequences. 
In a similar direction, the main obstacle for the existence of a full-fledged color/kinematics 
duality in the $\beta$-deformed theory is the appearance of symmetric structure constants.
The arguments of Leigh and Strassler suggest that it should not be possible to forbid their 
appearance without breaking conformal invariance at the quantum level. 
There may exist theories which, while exhibiting BCJ duality, also share some of the planar 
properties of the $\NeqFour$ theory. Identifying and studying such theories may shed further 
light on the relations between the symmetries  of the $\NeqFour$ theory, the possible 
patterns in which they can be broken and may also expose other unexpected features which 
are hidden by the large symmetry of this theory.

\section*{Acknowledgements:}

We would like to thank J.J.~Carrasco, L.~Dixon and H.~Johansson for useful discussions.
RR's work is supported in part by the US National Science Foundation under 
grant PHY-0855356 and the A. P. Sloan Foundation.

\end{document}